\documentclass[twocolumn,english,prb,citeautoscript,superscriptaddress]{revtex4-1}
\usepackage[T1]{fontenc}
\usepackage[latin9]{inputenc}
\usepackage{amsbsy}
\usepackage{amstext}
\usepackage{amssymb}
\usepackage{graphicx}
\usepackage{esint}

\makeatletter

\usepackage{amsfonts}
\usepackage{bbm}
\usepackage{graphics}
\newcommand{\upp}{\hspace{-0.2 pt}\uparrow}
\newcommand{\downn}{\hspace{-0.2 pt}\downarrow}

\def\ket#1{\left| #1\right>}
\def\bra#1{\left< #1\right|}

\usepackage{hyperref}
\hypersetup{colorlinks=true,citecolor=blue}
\def\<{\langle}
\def\>{\rangle}

\def\down{\downarrow}

\usepackage{leftidx}

\makeatother

\usepackage{babel}
\begin{document}

\title{Entangling distant resonant exchange qubits via circuit quantum electrodynamics}

\author{V. Srinivasa}

\email{vsriniv@umd.edu}

\selectlanguage{english}%

\affiliation{Laboratory for Physical Sciences, College Park, Maryland 20740, USA}

\affiliation{Department of Physics, University of Maryland, College Park, Maryland
20742, USA}

\author{J. M. Taylor}

\affiliation{Joint Quantum Institute, University of Maryland, College Park, Maryland
20742, USA}

\affiliation{Joint Center for Quantum Information and Computer Science, University
of Maryland, College Park, Maryland 20742, USA}

\affiliation{National Institute of Standards and Technology, Gaithersburg, Maryland,
20899, USA}

\author{C. Tahan}

\affiliation{Laboratory for Physical Sciences, College Park, Maryland 20740, USA}
\begin{abstract}
We investigate a hybrid quantum system consisting of spatially separated
resonant exchange qubits, defined in three-electron semiconductor
triple quantum dots, that are coupled via a superconducting transmission
line resonator. Drawing on methods from circuit quantum electrodynamics
and Hartmann-Hahn double resonance techniques, we analyze three specific
approaches for implementing resonator-mediated two-qubit entangling
gates in both dispersive and resonant regimes of interaction. We calculate
entangling gate fidelities as well as the rate of relaxation via phonons
for resonant exchange qubits in silicon triple dots and show that
such an implementation is particularly well-suited to achieving the
strong coupling regime. Our approach combines the favorable coherence
properties of encoded spin qubits in silicon with the rapid and robust
long-range entanglement provided by circuit QED systems. 
\end{abstract}
\maketitle

\section{Introduction }

The simultaneous realization of coherent local control of quantum
bits (qubits) and robust long-range interactions for entangling distant
qubits represents a fundamental goal in many proposed implementations
of quantum information processing \cite{DiVincenzo2000FortschrPhys,Ladd2010}.
Hybrid approaches may prove particularly advantageous for achieving
this goal, as they enable the optimal properties of multiple physically
distinct quantum systems to be combined \cite{Awschalom2013}. 

Electron spin qubits in semiconductor quantum dots promise protection
from environmental decoherence and potential scalability to larger
systems \cite{Loss1998,Taylor2005,Hanson2007RMP,Awschalom2013}. An
alternative to the challenges of coherently controlling single electron
spins via, e.g., highly localized magnetic fields \cite{Koppens2006},
spin-orbit coupling \cite{Nowack2007,Nadj-Perge2010}, or spin-position
coupling via magnetic field gradients \cite{Tokura2006,Pioro-Ladriere2008}
is provided by the encoding of single logical qubits in multiple electron
spins \cite{Levy2002,Petta2005,Taylor2005,DiVincenzo2000Nature,Laird2010,Medford2013NNano,Medford2013,Taylor2013}.
Within systems of two or more coupled quantum dots, such an encoding
effectively translates magnetic field control to electric field control,
providing a means of rapidly manipulating individual qubits, while
allowing for operation within decoherence-free subspaces that protect
against collective decoherence and leakage errors \cite{Lidar2000}. 

Three-electron spin qubits enable universal manipulation to be achieved
via electric field control of exchange interactions alone \cite{DiVincenzo2000Nature,Laird2010,Gaudreau2011,Medford2013NNano,Medford2013,Taylor2013,Shi2012,Kim2014,Eng2015}.
However, the electrostatic origin of exchange renders these spin qubits
susceptible to charge noise. The particular form of the exchange-only
qubit known as the resonant exchange (RX) qubit enables high-frequency
operation via resonant microwave driving of the exchange at a ``sweet
spot,'' where a large exchange energy gap suppresses the sensitivity
of the qubit to low-frequency charge noise \cite{Medford2013,Taylor2013,Fei2015,Russ2015,Shim2016arxiv}.
Thus, the RX qubit represents a semiconductor qubit similar to the
transmon superconducting qubit \cite{Koch2007,Houck2009} that additionally
possesses the protection intrinsic to spin qubits \cite{Hanson2007RMP,Awschalom2013}. 

A key element of a scalable quantum information processor based on
multielectron spin qubits is the implementation of long-range entangling
gates. The exchange interaction can be used to perform rapid, protected
gates between neighboring exchange-only qubits \cite{DiVincenzo2000Nature,Meier2003PRL,Meier2003PRB,Srinivasa2009,Fong2011,Doherty2013};
however, the range of the interaction is limited by its dependence
on the overlap of the wave functions of the electrons participating
in the coupling, which decreases exponentially with the separation
between electrons \cite{Burkard1999,Lidar2000}. Longer-range coupling
is possible via the Coulomb interaction and can be used to carry out
entangling gates between adjacent capacitively coupled exchange-only
qubits \cite{Taylor2013,Pal2014,Pal2015} without the leakage intrinsic
to two-qubit exchange gates \cite{DiVincenzo2000Nature,Meier2003PRL,Meier2003PRB,Fong2011,Doherty2013,Setiawan2014}.
The range of exchange-based interactions can be extended via spin
chains \cite{Friesen2007,Srinivasa2007}, while that of capacitive
interactions can be extended via floating metal gates \cite{Trifunovic2012}.
In the context of a modular quantum computer architecture \cite{Taylor2005,Monroe2014},
these schemes provide potential methods of coupling spatially separated
spin qubits within a single module.

In order to enable full scalability and modularity within a quantum
information processing device, however, a rapid and robust interaction
between qubits within different modules that are separated by macroscopic
distances is highly desirable \cite{Taylor2005,Monroe2014}. For superconducting
qubits, such an interaction can be realized within the approach of
circuit quantum electrodynamics (QED), where qubits separated by distances
of up to the order of centimeters are coupled capacitively to the
microwave field of a superconducting transmission line resonator \cite{Blais2004,Wallraff2004,Blais2007,Majer2007,Sillanpaa2007,DiCarlo2009,DiCarlo2010}.
An analogous approach has been investigated for hybrid solid-state
quantum systems consisting of semiconductor charge \cite{Childress2004,Frey2012,Delbecq2013,Toida2013,Liu2015,Gullans2015}
or spin \cite{Childress2004,Burkard2006,Taylor2006,Jin2012,Hu2012,Petersson2012,Tosi2014,Tosi2015arxiv}
qubits coupled to a superconducting resonator. The potentially longer
coherence times possible for spin qubits, particularly in silicon
\cite{Zwanenburg2013,Kawakami2014,Muhonen2014,Eng2015,Veldhorst2015,Takeda2016arxiv,Kawakami2016arxiv},
in comparison to superconducting qubits and quantum dot charge qubits
are advantageous for achieving the strong coupling regime, in which
the interaction rate exceeds both the qubit and cavity decay rates
\cite{Childress2004,Blais2004}. 

Here, we consider two RX qubits coupled via the fundamental mode of
a superconducting transmission line resonator as a potential building
block for a robust modular quantum information processing device.
The relatively large dipole moment of the RX qubit compared to double-dot
qubits, arising from the more delocalized three-electron wave function
within a triple quantum dot device \cite{Medford2013,Taylor2013},
should enable each RX qubit to interact strongly with the resonator
field. At the same time, the smaller size of the triple dot relative
to that of, e.g., a transmon \cite{DiCarlo2009,DiCarlo2010} should
prove useful in scaling to more qubits. Furthermore, the enhanced
protection of the RX qubit from low-frequency charge noise at its
optimal operating point in principle leads to coherence times sufficiently
long for realizing the strong coupling regime \cite{Taylor2013}. 

We begin by deriving the form of the coupling between a single RX
qubit and a transmission line resonator. Concurrent work investigates
this coupling in detail from a microscopic perspective \cite{Russ2015b}.
Subsequently, we analyze three specific approaches for resonator-mediated
entanglement of RX qubits based on techniques drawn from both circuit
QED and Hartmann-Hahn double resonance in NMR \cite{Hartmann1962}.
For each of these approaches, we explicitly derive the form of the
interaction and construct two-qubit entangling gates. Finally, we
explore the feasibility of implementing our proposed gates using RX
qubits defined within silicon triple quantum dots. We find that silicon
provides qubit relaxation times several orders of magnitude longer
than those for GaAs \cite{Taylor2013} and calculate gate fidelities
for the three approaches in the presence of qubit dephasing and cavity
decay. Our results suggest that the longer coherence times expected
for RX qubits in silicon enable the strong coupling regime to be attained
with currently achievable resonator quality factors and should lead
to high-fidelity two-qubit entangling gates in combination with rapid,
exchange-based universal single-qubit control \cite{Medford2013,Taylor2013}.

\section{\label{sec:RXrescoupling}Dipole coupling of a resonant exchange
qubit to a transmission line resonator}

\begin{figure}
\includegraphics[bb=10bp 100bp 550bp 480bp,width=3in]{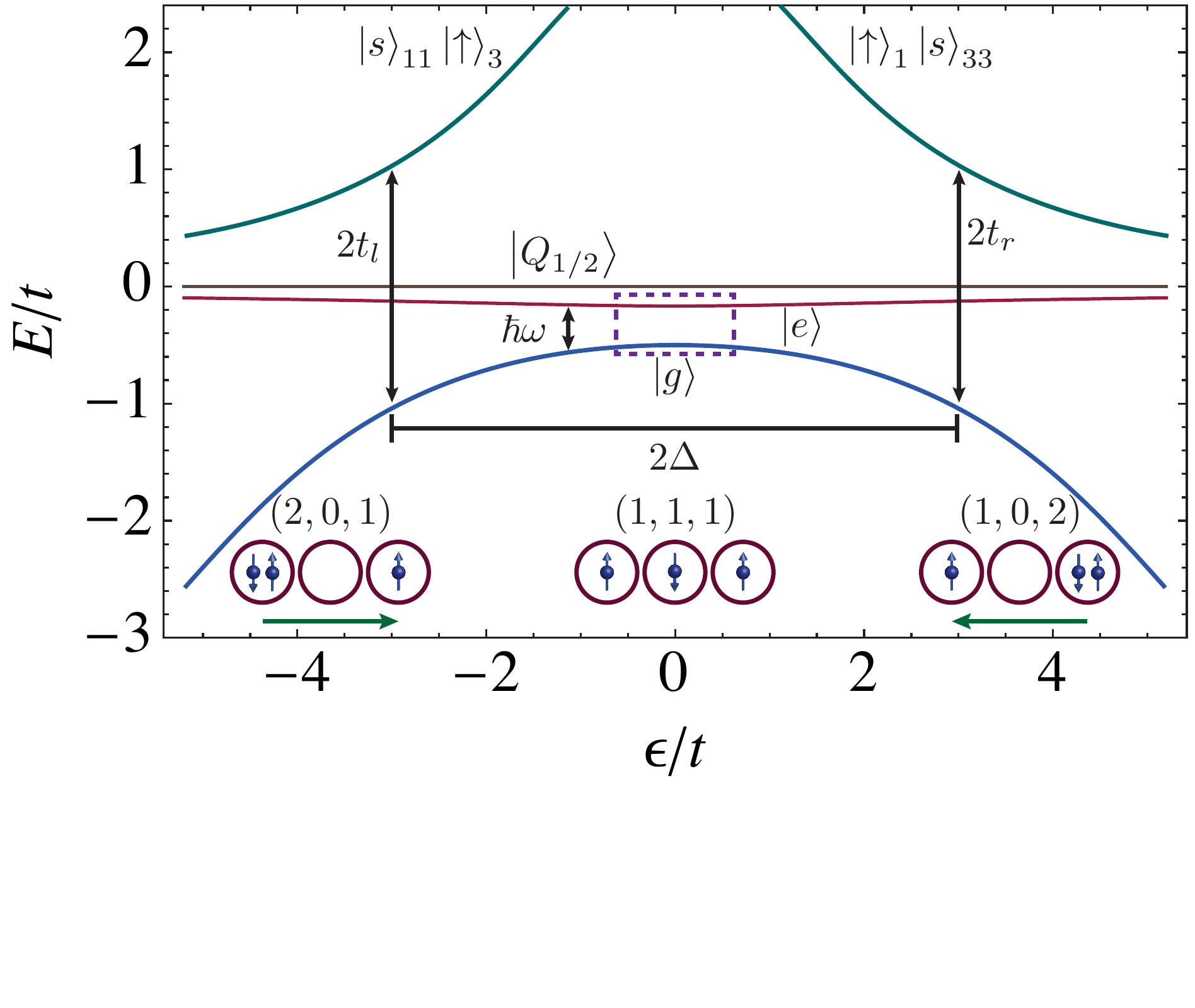}

\protect\caption{\label{fig:spectrum}Energy levels of a three-electron triple quantum
dot in the (1,1,1), (2,0,1), and (1,0,2) charge subspaces and the
$m_{s}=1/2$ spin subspace as a function of $\epsilon/t$ for $\Delta/t=3,$
with $t=t_{l}=t_{r}.$ A large uniform external magnetic field (for
which the Zeeman splitting is much larger than $t$) is assumed to
separate the subspace considered in the present work from other three-electron
states, and the remaining nearby state $\ket{Q_{1/2}}$ has $S=3/2$
\cite{Taylor2013}. The operating point considered in the present
work is indicated by a dotted rectangle. Insets illustrate schematically
the electric dipole moments of the triple dot in the (2,0,1) and (1,0,2)
configurations. }

\end{figure}

We consider a resonant exchange (RX) qubit, realized within a linear
triple quantum dot in the three-electron regime \cite{Medford2013,Taylor2013}.
In the present work, we follow Ref. \citenum{Taylor2013} and focus
on the subspace of three-electron states having total spin $S=1/2$
and spin quantum number for the total $z$ component $m_{s}=1/2$
in the charge subspaces (2,0,1), (1,1,1), and (1,0,2) (Fig. \ref{fig:spectrum}).
Here, $\left(n_{1},n_{2},n_{3}\right)$ denotes the charge occupation
of each dot (numbered from left to right). We assume that a uniform
external magnetic field is applied in the plane of the device \cite{Medford2013NNano}
in order to minimize its effect on the superconducting state of the
resonator \cite{Samkharadze2015arxiv} and is sufficiently large (typically
${\rm \gtrsim100\ mT}$ \cite{Medford2013}) such that other three-electron
states are energetically distant from the subspace we consider \cite{Taylor2013}.
For RX qubits implemented in silicon, we also assume that excited
valley states are well-separated in energy from the lowest-energy
valley manifold (see Sec. \ref{sec:ImplementSi}). The resulting subspace
is spanned by the (1,1,1) states 
\begin{eqnarray}
\ket{e_{0}} & \equiv & \ket{s}_{13}\ket{\upp}_{2}\nonumber \\
 & = & \frac{1}{\sqrt{2}}\left(c_{1\upp}^{\dagger}c_{2\upp}^{\dagger}c_{3\down}^{\dagger}-c_{1\downn}^{\dagger}c_{2\upp}^{\dagger}c_{3\upp}^{\dagger}\right)\ket{\mathrm{vac}},\label{eq:state1}\\
\ket{g_{0}} & \equiv & \sqrt{\frac{2}{3}}\ket{t_{+}}_{13}\ket{\downn}_{2}-\sqrt{\frac{1}{3}}\ket{t_{0}}_{13}\ket{\upp}_{2}\nonumber \\
 & = & \frac{1}{\sqrt{6}}\left(2c_{1\upp}^{\dagger}c_{2\downn}^{\dagger}c_{3\upp}^{\dagger}-c_{1\upp}^{\dagger}c_{2\upp}^{\dagger}c_{3\down}^{\dagger}-c_{1\downn}^{\dagger}c_{2\upp}^{\dagger}c_{3\upp}^{\dagger}\right)\ket{\mathrm{vac}},\nonumber \\
\label{eq:state0}
\end{eqnarray}
along with the (2,0,1) and (1,0,2) states
\begin{eqnarray}
\ket{s_{1,1/2}} & \equiv & \ket{s}_{11}\ket{\uparrow}_{3}=c_{1\uparrow}^{\dagger}c_{1\downarrow}^{\dagger}c_{3\uparrow}^{\dagger}\ket{\mathrm{vac}},\label{eq:stateS11}\\
\ket{s_{3,1/2}} & \equiv & \ket{\uparrow}_{1}\ket{s}_{33}=c_{1\uparrow}^{\dagger}c_{3\uparrow}^{\dagger}c_{3\downarrow}^{\dagger}\ket{\mathrm{vac}},\label{stateS33}
\end{eqnarray}
where $c_{i\sigma}^{\dagger}$ is the creation operator for an electron
in dot $i$ with spin $\sigma,$ and $\ket{\rm vac}$ denotes the
vacuum. 

When restricted to the (1,1,1) charge subspace, these states also
define the originally proposed exchange-only qubit \cite{DiVincenzo2000Nature}.
As discussed in detail in Ref. \citenum{Taylor2013}, the RX qubit
is defined within an effective (1,1,1) subspace, spanned by the states
$\tilde{\ket{g_{0}}}$ and $\tilde{\ket{e_{0}}}$ obtained via elimination
of the higher-energy three-electron states with charge configurations
(2,0,1) and (1,0,2) in Eqs. (\ref{eq:stateS11}) and (\ref{stateS33})
using a Schrieffer-Wolff transformation. Defining $\tilde{\sigma}^{z}\equiv\tilde{\ket{e_{0}}}\tilde{\bra{e_{0}}}-\tilde{\ket{g_{0}}}\tilde{\bra{g_{0}}},$
the resulting effective Hamiltonian in this subspace is given by 
\begin{equation}
H_{{\rm eff}}=\frac{J}{2}\tilde{\sigma}^{z}-\frac{\sqrt{3}}{2}j\tilde{\sigma}^{x},\label{eq:Heff}
\end{equation}
with $J\equiv\left(J_{l}+J_{r}\right)/2,$ $j\equiv\left(J_{l}-J_{r}\right)/2,$
and where the exchange between the center and left (right) dots is
$J_{l}=t_{l}^{2}/\left(\Delta+\epsilon\right)$ $\left[J_{r}=t_{r}^{2}/\left(\Delta-\epsilon\right)\right].$
Here, $t_{l}$ and $t_{r}$ are the corresponding tunneling amplitudes,
$\epsilon\equiv\left(V_{3}-V_{1}\right)/2$ is the detuning parameter
for the RX qubit with on-site energies $-V_{1}$ and $-V_{3}$ for
the left and right dots, respectively, and $\Delta$ is defined in
terms of Hubbard model parameters such that $\Delta+\epsilon$ $\left(\Delta-\epsilon\right)$
is the energy of the state $\ket{s_{1,1/2}}$ $\left(\ket{s_{3,1/2}}\right)$
relative to that of the states $\ket{g_{0}}$ and $\ket{e_{0}}$ (see
Ref. \citenum{Taylor2013} for more details). In Eq. (\ref{eq:Heff})
and the expressions given throughout the remainder of this work, we
neglect terms proportional to the identity operator unless otherwise
noted. Importantly, while the uniform charge state (1,1,1) of a three-electron
triple dot does not interact directly with the microwave field of
the resonator, the logical states of the RX qubit contain small admixtures
of the polarized charge states (2,0,1) and (1,0,2), enabling coupling
to the resonator field (Fig. \ref{fig:spectrum}). 

In order to determine the strength of this coupling, we first consider
the electric dipole transition matrix element for the RX qubit. We
take the centers of the three dots to be located at positions $\mathbf{r}_{1}=-\left(w/2\right)\hat{x},$
$\mathbf{r}_{2}=\mathbf{0},$ and $\mathbf{r}_{3}=\left(w/2\right)\hat{x}.$
The dipole operator for the triple dot can then be written as $\mathbf{d}=-e\sum_{j}\mathbf{r}_{j}n_{j}\equiv d\hat{x},$
where $d=\frac{ew}{2}\left(n_{1}-n_{3}\right)$ and $e$ is the magnitude
of the electron charge. In the chosen subspace $\left\{ \ket{e_{0}},\ket{g_{0}},\ket{s_{1,1/2}},\ket{s_{3,1/2}}\right\} $,
the dipole operator takes the form 
\begin{eqnarray}
d & = & \frac{ew}{2}\left(\left|s_{1,1/2}\right\rangle \left\langle s_{1,1/2}\right|-\left|s_{3,1/2}\right\rangle \left\langle s_{3,1/2}\right|\right).\label{eq:d}
\end{eqnarray}
Applying the same Schrieffer-Wolff transformation used to obtain $H_{{\rm eff}}$
{[}Eq. (\ref{eq:Heff}){]} then yields 
\begin{equation}
\tilde{d}=\frac{ew}{2}\left(\frac{1}{2}\partial_{\epsilon}J\tilde{\sigma}^{z}-\frac{\sqrt{3}}{2}\partial_{\epsilon}j\tilde{\sigma}_{}^{x}\right)=\frac{ew}{2}\partial_{\epsilon}H_{{\rm eff}}.\label{eq:dtilde}
\end{equation}

Introducing a small variation $F$ in the detuning such that $\epsilon=\epsilon_{0}+F,$
we write the Hamiltonian of the RX qubit as $H\approx\left.H_{{\rm eff}}\right|_{\epsilon=\epsilon_{0}}+\left.\partial_{\epsilon}H_{{\rm eff}}\right|_{\epsilon=\epsilon_{0}}F.$
Note that by comparing this expansion with the expression for $\tilde{d}$
in Eq. (\ref{eq:dtilde}), the second term of $H$ can be identified
with the standard dipole coupling Hamiltonian. The operating point
$\epsilon=\epsilon_{0}$ for the RX qubit is chosen such that the
coupling to $F,$ which is proportional to $\partial_{\epsilon}H_{{\rm eff}},$
is perpendicular to the quantization axis in a Bloch sphere representation
of the RX qubit. This choice corresponds to the condition $\partial_{\epsilon}J=-\left(3j/J\right)\partial_{\epsilon}j.$
Defining $\sigma^{z}\equiv\left|e\right\rangle \left\langle e\right|-\left|g\right\rangle \left\langle g\right|,$
where $\left\{ \ket{g},\ket{e}\right\} $ is the basis that diagonalizes
$H_{{\rm eff}}$ {[}Eq. (\ref{eq:Heff}){]}, the Hamiltonian for small
variations of the detuning about the operating point $\epsilon=\epsilon_{0}$
is given by 
\begin{equation}
H_{{\rm RX}}=\frac{\hbar\omega}{2}\sigma^{z}+F\eta\sigma^{x},\label{eq:Hrx}
\end{equation}
 where $\hbar\omega\equiv\sqrt{J^{2}+3j^{2}}$ and $\eta\equiv\left.\frac{1}{2}\sqrt{\left(\partial_{\epsilon}J\right)^{2}+3\left(\partial_{\epsilon}j\right)^{2}}\right|_{\epsilon=\epsilon_{0}}.$
Thus, the effective dipole transition matrix element for the RX qubit
is given by 
\begin{eqnarray}
\bra{g}d\ket{e} & = & \frac{ew}{2}\eta=\left.\frac{ew}{4}\sqrt{\left(\partial_{\epsilon}J\right)^{2}+3\left(\partial_{\epsilon}j\right)^{2}}\right|_{\epsilon=\epsilon_{0}}.\label{eq:Hrx01}
\end{eqnarray}

\begin{figure}
\includegraphics[bb=0bp 0bp 420bp 253bp,width=3.2in]{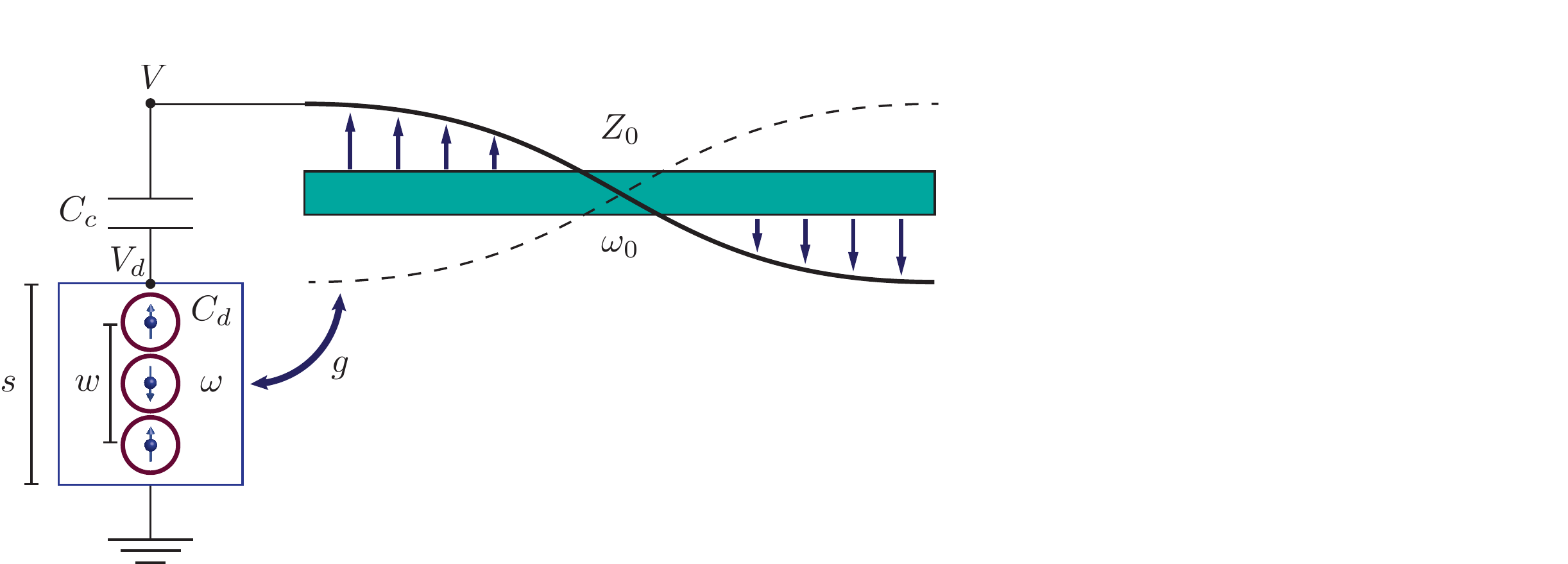}

\protect\caption{\label{fig:RXrescoupling}Schematic illustration of the geometry and
parameters used to determine the RX qubit-resonator coupling strength
$g$. For clarity, only the center conductor of the resonator is shown. }
\end{figure}

We now determine the coupling strength for the interaction of the
RX qubit with the lowest-energy mode of a transmission line resonator
(Fig. \ref{fig:RXrescoupling}). A resonator of length $l,$ capacitance
per unit length $C_{0},$ and inductance per unit length $L_{0}$
has a characteristic impedance $Z_{0}=\sqrt{L_{0}/C_{0}}$. The fundamental
mode of the resonator has frequency $\omega_{0}=\pi/lZ_{0}C_{0}$
and is described by the quantized antinode voltage \cite{Childress2004,Taylor2006}
\begin{equation}
\hat{V}=\sqrt{\frac{\hbar\omega_{0}}{lC_{0}}}\left(a+a^{\dagger}\right).\label{eq:Vtlr}
\end{equation}
where $a$ and $a^{\dagger}$ are resonator photon creation and annihilation
operators. We assume a geometry of the type shown in Fig. \ref{fig:RXrescoupling}
for the capacitive coupling between the resonator and the triple dot,
with the antinodes of the resonator field positioned at the two ends
of the center conductor of the resonator. In terms of the total capacitance
$C_{c}$ between the resonator and the triple dot and the total capacitance
$C_{d}$ between the triple dot and ground, the voltage across the
triple dot is given by $\hat{V}_{d}=C_{c}\hat{V}/\left(C_{c}+C_{d}\right)\equiv v\hat{V}.$
The interaction of the RX qubit with this voltage, described by $H_{{\rm int}}=-{\bf d}\cdot{\bf E}=d\hat{V}_{d}/s,$
where $s$ is the effective distance over which the voltage drop $\hat{V}_{d}$
occurs, then takes the form
\begin{equation}
H_{{\rm int}}=\hbar g_{0}\left(n_{1}-n_{3}\right)\left(a+a^{\dagger}\right),\label{eq:Hint}
\end{equation}
with the vacuum Rabi coupling
\begin{equation}
g_{0}\equiv\frac{ewv}{2slC_{0}}\sqrt{\frac{\pi}{Z_{0}\hbar}}=\frac{ewv}{2s}\omega_{0}\sqrt{\frac{Z_{0}}{\pi\hbar}}.\label{eq:g0}
\end{equation}
In the representation $\left\{ \tilde{\ket{e_{0}}},\tilde{\ket{g_{0}}}\right\} $,
the operator $\left(n_{1}-n_{3}\right)$ becomes $\partial_{\epsilon}H_{{\rm eff}}.$
Since the dipole coupling Hamiltonian $\tilde{H}_{{\rm int}}=\left.\partial_{\epsilon}H_{{\rm eff}}\right|_{\epsilon=\epsilon_{0}}F$
as discussed above, we also note that the interaction with the transmission
line resonator is described by quantizing $F$ and setting $F=ew\hat{V}_{d}/2s$
so that the oscillation in the detuning of the RX qubit is controlled
by the resonator voltage. At the operating point $\epsilon=\epsilon_{0},$
which corresponds to $\partial_{\epsilon}J=-\left(3j/J\right)\partial_{\epsilon}j,$
the interaction Hamiltonian in the qubit basis $\left\{ \ket{g},\ket{e}\right\} $
takes the form
\begin{equation}
H_{{\rm int}}^{\prime}=\hbar g\sigma^{x}\left(a+a^{\dagger}\right),\label{eq:Hintpr}
\end{equation}
where the effective qubit-resonator coupling strength is given by
\begin{eqnarray}
g & \equiv & \eta g_{0}\nonumber \\
 & = & \left.\frac{g_{0}}{2}\sqrt{\left(\partial_{\epsilon}J\right)^{2}+3\left(\partial_{\epsilon}j\right)^{2}}\right|_{\epsilon=\epsilon_{0}}.\label{eq:geff}
\end{eqnarray}

To estimate the interaction strength $g,$ we choose $\epsilon_{0}=0$
as the operating point for the qubit (see Fig. \ref{fig:spectrum}).
At this point, $J_{l}=J_{r}=t^{2}/\Delta,$ while $\partial_{\epsilon}J=0$
and $\partial_{\epsilon}j=-t^{2}/\Delta^{2}.$ Defining the charge
admixture parameter $\xi\equiv t/\Delta,$ this gives $\eta=\sqrt{3}\xi^{2}/2.$
For simplicity, we set $s=w$ in Eq. (\ref{eq:g0}). Choosing $Z_{0}=50\ \Omega,$
$v=0.28$ \cite{Childress2004,Blais2004}, and $\omega_{0}=2\pi\times1.5\ {\rm GHz}$
\cite{Taylor2006} (where we assume $k_{B}T\ll\hbar\omega_{0}$ in
order to neglect thermal excitation of photons) leads to a charge-cavity
coupling strength $g_{0}\approx2\pi\times13\ {\rm MHz}.$ Taking $\xi=0.3$
\cite{Taylor2013}, we then find $g\approx2\pi\times1\ {\rm MHz}$
for the strength of the coupling between the RX qubit and the transmission
line resonator, which is comparable to that found experimentally for
spin qubits in double quantum dots coupled to a higher-frequency resonator
\cite{Petersson2012}. We note that, while the qubit-resonator coupling
strength $g$ is limited by the constraint $\xi\ll1$ required for
the validity of the Heisenberg model description of the RX qubit \cite{Taylor2013},
the estimate we obtain here does not represent a fundamental limit
and varies strongly with the chosen parameters. Further enhancements
of $g$ may be possible by, e.g., modifying the design of the transmission
line resonator to increase the strength of the electric field within
the region containing the RX qubit \cite{Tosi2014,Samkharadze2015arxiv}
and thus the charge-cavity coupling strength $g_{0}.$ Recent work
\cite{Samkharadze2015arxiv} demonstrates a characteristic impedance
$Z_{0}\approx4\ {\rm k}\Omega$ for superconducting nanowire resonators
with high kinetic inductance, which leads to $g_{0}\approx2\pi\times120\ {\rm MHz}$
and $g\approx2\pi\times9\ {\rm MHz}$ for the same values of $\omega_{0},$
$v$ and $\xi$ chosen above. 

The value of $g$ we determine here is nevertheless sufficient for
reaching the strong coupling regime of the RX qubit-resonator interaction
in realistic systems, as we now show. The strong coupling regime corresponds
to $g>\kappa,\gamma,$ where $\kappa=\omega_{0}/Q$ is the rate of
decay of microwave photons out of the resonator, $Q$ is the resonator
quality factor, and $\gamma=1/T_{2}^{\ast}$ is the qubit decay rate
\cite{Childress2004,Blais2004}. This regime is characterized by a
transfer of excitations between the qubit and the resonator that is
more rapid than the qubit and photon decay rates and is thus of fundamental
importance for resonator-mediated entangling gates. For the estimated
coupling strength $g=2\pi\times9\ {\rm MHz}$ and $\omega_{0}=2\pi\times1.5\ {\rm GHz},$
the strong coupling regime can be achieved when the conditions $Q>\omega_{0}/g\approx160$
and $T_{2}^{\ast}>1/g\approx17\ {\rm ns}$ are satisfied. Recent experiments
\cite{Frey2012,Petersson2012,Liu2015} suggest that sufficiently high
quality factors are already attainable in coupled dot-resonator systems.
In addition, a single RX qubit implemented in a GaAs triple quantum
dot has a dephasing time $T_{2}^{\ast}\approx500\ n{\rm s},$ and
an increase of the coherence time to $T_{2}\approx20\ \mu{\rm s}$
via echo has been demonstrated \cite{Medford2013}. The strong coupling
regime should therefore be attainable for $\xi\ll0.3,$ which is advantageous
for RX qubits in GaAs dots as the rate of qubit decay due to phonons
is expected to increase as $\sim\xi^{4}$ \cite{Taylor2013}. In Sec.
\ref{sec:ImplementSi}, we consider relaxation due to electron-phonon
coupling for RX qubits in Si triple dots and find that, for this system,
phonon-induced decay is unlikely to limit qubit coherence. Thus, Si-based
RX qubits should in principle enable operation in the strong coupling
regime for larger dipole moments (i.e., larger $\eta\sim\xi^{2}$),
which should lead to more rapid entangling gates. 

\begin{figure}
\includegraphics[bb=10bp 20bp 550bp 176bp,width=3.2in]{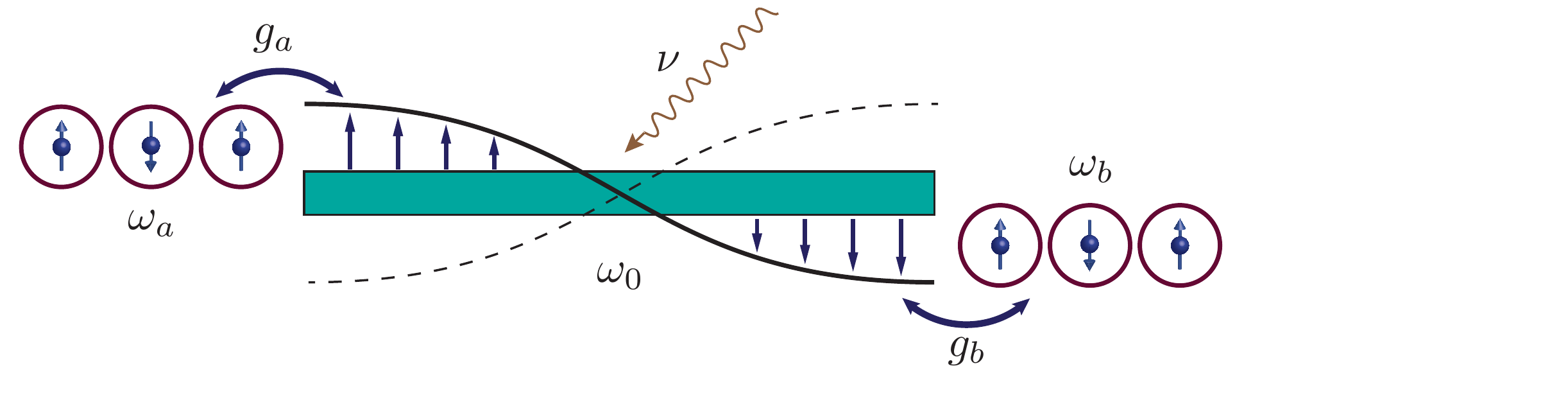}

\protect\caption{\label{fig:RXresRX}Schematic showing two resonant exchange (RX) qubits,
having transition frequencies $\omega_{a}$ and $\omega_{b},$ coupled
to the fundamental mode of a microwave transmission line resonator,
having frequency $\omega_{0},$ with strengths $g_{a}$ and $g_{b},$
respectively. An external microwave driving field of frequency $\nu$
applied to the resonator is also indicated. }
\end{figure}

\section{\label{sec:RXresRX}Resonator-mediated entangling gates for two rx
qubits}

We now explore three specific approaches for entangling two spatially
separated RX qubits which are both coupled to the fundamental mode
of a transmission line resonator via their electric dipole moments
(Fig. \ref{fig:RXresRX}). As we show below, methods developed for
coupling superconducting qubits in circuit QED \cite{Blais2004,Blais2007}
along with Hartmann-Hahn double resonance techniques for externally
driven systems \cite{Hartmann1962} can be directly applied to RX
qubits, in combination with fast single-qubit rotations via exchange.
We explicitly derive the effective interactions and the two-qubit
entangling gates they generate for each approach, while including
relevant prior results from circuit QED for completeness. 

In Sec. \ref{sub:disp}, we consider the interaction between two RX
qubits with identical transition frequencies mediated by virtual excitations
of the resonator in the dispersive regime, which gives rise to gate
rates $\sim g^{2}/\tilde{\Delta}\sim\xi^{4},$ where $\tilde{\Delta}$
denotes the detuning of the qubit transition frequency with respect
to the resonator frequency. We describe an alternative approach in
Sec. \ref{sub:resonant}, which involves driving each qubit with resonant
microwave fields to generate sideband transitions in a doubly rotating
frame. This approach leads to a gate rate that scales linearly with
the coupling strength $g\sim\xi^{2},$ potentially providing a faster
gate than that discussed in Sec. \ref{sub:disp}. Finally, we consider
the dispersive regime for the driven resonator-mediated interaction
between two RX qubits having different transition frequencies in Sec.
\ref{sub:drivdisp}. We show how transforming to a doubly rotating
frame enables two-qubit entangling gates with rates $\sim\xi^{4}$
even in the presence of the qubit frequency variation. 

Identifying the regime which optimizes two-qubit gate fidelities for
a particular implementation involves a comparison of the qubit and
resonator photon coherence times. Two-qubit gates in the dispersive
regime are expected to have higher fidelities for systems where the
qubit coherence time exceeds the resonator photon coherence time,
as the interaction is mediated only by virtual resonator photons.
On the other hand, fidelities for two-qubit gates based on the direct
qubit-resonator interaction in the resonant regime are limited by
both the qubit and the resonator photon coherence times. In Sec. \ref{sec:EntanglingF},
we show that this intuition is consistent with calculated gate fidelities
for the three regimes we consider.

\subsection{\label{sub:disp}Dispersive regime}

Initially, we consider two RX qubits coupled to the resonator in the
absence of external driving fields. Using Eqs. (\ref{eq:Hrx}) and
(\ref{eq:Hintpr}), the Hamiltonian of the combined system can be
written as $H_{d}=H_{0}+V,$ where (setting $\hbar=1$) 
\begin{equation}
H_{0}=\omega_{0}a^{\dagger}a+\sum_{\mu=a,b}\frac{\omega_{\mu}}{2}\sigma_{\mu}^{z}\label{eq:H0}
\end{equation}
describes the individual RX qubits (having transition frequencies
$\omega_{a}$ and $\omega_{b}$) together with the fundamental mode
of the resonator (having frequency $\omega_{0}$), and 
\begin{equation}
V=\sum_{\mu=a,b}g_{\mu}\sigma_{\mu}^{x}\left(a+a^{\dagger}\right)\label{eq:V}
\end{equation}
describes the dipolar interaction of the RX qubits with the resonator.
Thus, $H_{d}$ is of the Jaynes-Cummings form \cite{Meystre2007}
with $g_{\mu}\equiv\eta_{\mu}g_{0}^{\mu}$ representing the strength
of the coupling to the resonator for qubit $\mu=a,b.$ By using $\sigma_{\mu}^{x}=\sigma_{\mu}^{+}+\sigma_{\mu}^{-},$
where $\sigma_{\mu}^{+}\equiv\ket{e}\bra{g}$ and $\sigma_{-}^{\mu}\equiv\ket{g}\bra{e},$
defining the qubit-resonator detunings $\tilde{\Delta}_{\mu}\equiv\omega_{0}-\omega_{\mu},$
and applying a rotating wave approximation for $\left|\tilde{\Delta}_{\mu}\right|\ll\omega_{0}+\omega_{\mu}$
in order to neglect the energy-nonconserving terms containing $\sigma_{\mu}^{+}a^{\dagger}$
and $\sigma_{\mu}^{-}a,$ Eq. (\ref{eq:V}) can be approximated as
\begin{equation}
V\approx\sum_{\mu=a,b}g_{\mu}\left(\sigma_{\mu}^{+}a+\sigma_{\mu}^{-}a^{\dagger}\right).\label{eq:Vrwa}
\end{equation}

The effective two-qubit interaction mediated by the resonator between
the RX qubits in the dispersive regime, defined by $g_{\mu}\ll\left|\tilde{\Delta}_{\mu}\right|$
for $\mu=a,b,$ is obtained via a Schrieffer-Wolff transformation
that eliminates the direct qubit-resonator coupling {[}Eq. (\ref{eq:Vrwa}){]}
to first order in $g_{\mu}/\tilde{\Delta}_{\mu}$ \cite{Blais2004,Burkard2006,Blais2007}.
The resulting effective Hamiltonian can be approximated as $\tilde{H}_{d}\equiv H_{0}+\frac{1}{2}\left[S_{1},V\right],$
where
\begin{equation}
S_{1}\equiv-\sum_{\mu=a,b}\frac{g_{\mu}}{\tilde{\Delta}_{\mu}}\left(\sigma_{\mu}^{+}a-\sigma_{\mu}^{-}a^{\dagger}\right).\label{eq:S1}
\end{equation}
We express the result in the form $\tilde{H}_{d}=\tilde{H}_{0}+\tilde{V},$
with

\begin{eqnarray}
\tilde{H}_{0} & \equiv & H_{0}-\sum_{\mu=a,b}\frac{g_{\mu}^{2}}{\tilde{\Delta}_{\mu}}\left(a^{\dagger}a+\frac{1}{2}\right)\sigma_{\mu}^{z},\label{eq:H0tilde}\\
\tilde{V} & \equiv-\chi_{ab} & \left(\sigma_{a}^{+}\sigma_{b}^{-}+\sigma_{a}^{-}\sigma_{b}^{+}\right),\label{eq:Vtilde}
\end{eqnarray}
where we define the two-qubit coupling strength
\begin{equation}
\chi_{ab}\equiv\frac{g_{a}g_{b}}{2}\left(\frac{1}{\tilde{\Delta}_{a}}+\frac{1}{\tilde{\Delta}_{b}}\right).\label{eq:chiab}
\end{equation}

To obtain the two-qubit entangling gate generated by $\tilde{H}_{d},$
we focus on the zero-photon two-qubit subspace $\left\{ \ket{e,e,0},\ket{e,g,0},\ket{g,e,0},\ket{g,g,0}\right\} $.
Here, $\ket{n}$ denotes the \emph{n}-photon state of the resonator,
with $n=0,1,2,\ldots.$ In this subspace, we find 
\begin{eqnarray}
\tilde{H}_{d}^{\left(0\right)} & \equiv & \bra{0}\tilde{H}_{d}\ket{0}\nonumber \\
 & = & \sum_{\mu=a,b}\frac{\omega_{\mu}^{\prime}}{2}\sigma_{\mu}^{z}\nonumber \\
 & - & \chi_{ab}\left(\sigma_{a}^{+}\sigma_{b}^{-}+\sigma_{a}^{-}\sigma_{b}^{+}\right).\label{eq:Hdtilde0}
\end{eqnarray}
Eq. (\ref{eq:Hdtilde0}) is expressed in terms of the modified qubit
transition frequency 
\begin{equation}
\omega_{\mu}^{\prime}\equiv\omega_{\mu}-\frac{g_{\mu}^{2}}{\tilde{\Delta}_{\mu}}.\label{eq:omegamuprime}
\end{equation}
We now transform $\tilde{H}_{d}^{\left(0\right)}$ to a rotating frame
via 
\begin{equation}
U_{{\rm rf}}=e^{-i(\omega_{a}^{\prime}\sigma_{a}^{z}+\omega_{b}^{\prime}\sigma_{b}^{z})\tau/2}.\label{eq:Urf}
\end{equation}
This leads to 
\begin{eqnarray}
H_{d}^{{\rm rf}} & = & U_{{\rm rf}}^{\dagger}\tilde{H}_{d}^{\left(0\right)}U_{{\rm rf}}-iU_{{\rm rf}}^{\dagger}\dot{U}_{{\rm rf}}\nonumber \\
 & = & -\chi_{ab}\left[\sigma_{a}^{+}\sigma_{b}^{-}e^{i\left(\omega_{a}^{\prime}-\omega_{b}^{\prime}\right)\tau}+\sigma_{a}^{-}\sigma_{b}^{+}e^{-i\left(\omega_{a}^{\prime}-\omega_{b}^{\prime}\right)\tau}\right].\label{eq:Hdrf}
\end{eqnarray}
Note that both $U_{d}$ and $H_{d}^{{\rm rf}}$ act only within the
zero-photon subspace. When the transition frequencies of the two qubits
are resonant with each other, so that $\omega_{a}^{\prime}=\omega_{b}^{\prime},$
we find 
\begin{eqnarray}
H_{d}^{{\rm rf}} & \approx & -\frac{g_{a}g_{b}}{\tilde{\Delta}}\left(\sigma_{a}^{+}\sigma_{b}^{-}+\sigma_{a}^{-}\sigma_{b}^{+}\right),\label{eq:Hdrfapprox}
\end{eqnarray}
where we have dropped small terms $\sim g_{\mu}^{2}/\tilde{\Delta}_{\mu}^{2}$
so that $\tilde{\Delta}_{a}^{-1}\approx\tilde{\Delta}_{b}^{-1}\equiv\tilde{\Delta}^{-1}.$ 

The unitary evolution generated by the interaction in Eq. (\ref{eq:Hdrfapprox})
is described by the operator \cite{Blais2004} 
\begin{eqnarray}
U_{d}\left(\tau\right) & \equiv & e^{-iH_{d}^{{\rm rf}}\tau}\nonumber \\
 & = & \left(\begin{array}{cccc}
1\\
 & \cos\left(\frac{g_{a}g_{b}\tau}{\tilde{\Delta}}\right) & i\sin\left(\frac{g_{a}g_{b}\tau}{\tilde{\Delta}}\right)\\
 & i\sin\left(\frac{g_{a}g_{b}\tau}{\tilde{\Delta}}\right) & \cos\left(\frac{g_{a}g_{b}\tau}{\tilde{\Delta}}\right)\\
 &  &  & 1
\end{array}\right).\nonumber \\
\label{eq:Ud}
\end{eqnarray}
For $\tau=\tau_{n}\equiv\left(4n+1\right)\pi\tilde{\Delta}/2g_{a}g_{b},$
where $n$ is an integer, Eq. (\ref{eq:Ud}) yields the \emph{$i{\rm SWAP}$}
gate. This two-qubit entangling gate can be combined with single qubit
rotations to form a universal set of quantum gates \cite{Schuch2003}.
The gate $U_{d}$ has a rate given by $g_{a}g_{b}/\tilde{\Delta}\sim g^{2}/\tilde{\Delta}\sim\xi^{4},$
where we assume $g_{b}\sim g_{a}\equiv g.$

\subsection{\label{sub:resonant}Driven resonant regime}

The two-qubit entangling gates considered in Sec. \ref{sub:disp}
have rates that scale as $g^{2}/\tilde{\Delta}\sim\xi^{4}.$ Thus,
these gates are limited in speed by both the condition $\xi\ll1$
for the validity of the Heisenberg model \cite{Taylor2013} and the
fact that they are carried out in the dispersive regime, where $g\ll\tilde{\Delta}.$
In order to obtain more rapid gates with rates that vary linearly
with $g\sim\xi^{2},$ we now consider an alternative approach for
generating entanglement between two RX qubits based on microwave driving
of sideband transitions \cite{Blais2007,Wallraff2007,Leek2009}, in
the spirit of the Cirac-Zoller gate for two-level ions \cite{Cirac1995,Childs2000}.
When a microwave driving field resonant with one of the qubit transitions
is applied to the resonator, the qubit-resonator interaction {[}Eq.
(\ref{eq:Hintpr}){]} enables driving of the qubit at its Rabi frequency.
In a frame rotating at both the drive and Rabi frequencies, the interaction
of the resonator with each qubit leads to sideband transitions for
appropriately chosen frequencies of the driving field \cite{Blais2007}.
These transitions can be used to construct two-qubit entangling gates
\cite{Cirac1995,Childs2000}. Here, we show how this approach applies
directly to two RX qubits coupled to a transmission line resonator
and explicitly derive the full gate sequence for a controlled-Z ($\pi$-phase)
gate. 

To obtain the effective Hamiltonians which generate sideband transitions,
we begin by considering the interaction of a single RX qubit with
the resonator in the presence of an external driving field. Writing
Eqs. (\ref{eq:H0}) and (\ref{eq:V}) for a single qubit and adding
a term describing driving of the resonator (see Fig. \ref{fig:RXresRX})
by an applied microwave field of frequency $\nu,$ amplitude $\varepsilon,$
and phase $\phi$ gives 
\begin{eqnarray}
H_{r} & = & \omega_{0}\tilde{a}^{\dagger}\tilde{a}+\frac{\omega}{2}\sigma^{z}+g\sigma^{x}\left(\tilde{a}+\tilde{a}^{\dagger}\right)\nonumber \\
 & + & \varepsilon\left[e^{-i\left(\nu\tau+\phi\right)}\tilde{a}^{\dagger}+e^{i\left(\nu\tau+\phi\right)}\tilde{a}\right],\label{eq:Hr}
\end{eqnarray}
where we use a modified notation for the resonator mode operators
$\tilde{a}$ and $\tilde{a}^{\dagger}$ for convenience in later expressions.
We work in a regime where the driving field amplitude $\varepsilon$
and the detuning $\Delta_{0}\equiv\omega_{0}-\nu$ are sufficiently
large that we can neglect the quantum fluctuations of the driving
field arising from its interaction with the resonator and approximate
the drive as a classical field \cite{Blais2007}. In this regime,
we can eliminate terms describing the direct action of the drive on
the resonator from the Hamiltonian by applying a displacement transformation
using $D\left(\alpha\right)\equiv e^{\alpha\tilde{a}^{\dagger}-\alpha^{\ast}\tilde{a}}$
and setting $\alpha\left(\tau\right)$ equal to the steady-state solution
of $\dot{\alpha}+i\omega_{0}\alpha+i\varepsilon e^{-i(\nu\tau+\phi)}=0,$
which gives $\alpha\left(\tau\right)=-\varepsilon e^{-i\left(\nu\tau+\phi\right)}/\Delta_{0}.$
This choice of $\alpha$ yields

\begin{eqnarray}
H_{r}^{\prime} & = & D^{\dagger}\left(\alpha\right)H_{r}D\left(\alpha\right)-iD^{\dagger}\left(\alpha\right)\dot{D}\left(\alpha\right)\nonumber \\
 & = & \omega_{0}\tilde{a}^{\dagger}\tilde{a}+\frac{\omega}{2}\sigma^{z}+g\sigma^{x}\left(\tilde{a}+\tilde{a}^{\dagger}\right)\nonumber \\
 & - & 2\Omega\cos\left(\nu\tau+\phi\right)\sigma^{x},\label{eq:Hrprime}
\end{eqnarray}
where 
\begin{eqnarray}
2\Omega & \equiv & \frac{2g\varepsilon}{\Delta_{0}}\label{eq:Rabifreq}
\end{eqnarray}
is the Rabi frequency for the qubit.

We next transform to a frame rotating at the drive frequency $\nu$
using the unitary transformation 
\begin{equation}
U_{1}=e^{-i\nu\tau\left(\tilde{a}^{\dagger}\tilde{a}+\sigma^{z}/2\right)}.\label{eq:U1}
\end{equation}
Driving the qubit on resonance, so that $\nu=\omega,$ and dropping
rapidly oscillating terms $\sim e^{\pm2i\nu\tau}$ leads to 
\begin{eqnarray}
H_{r}^{{\rm rf}} & = & \Delta_{0}\tilde{a}^{\dagger}\tilde{a}+g\left(\sigma^{+}\tilde{a}+\sigma^{-}\tilde{a}^{\dagger}\right)\nonumber \\
 & - & \Omega\left(e^{-i\phi}\sigma^{+}+e^{i\phi}\sigma^{-}\right).\label{eq:Hrrf}
\end{eqnarray}
For convenience, we rotate $H_{r}^{{\rm rf}}$ such that the last
term becomes proportional to $\sigma^{y}.$ This can be achieved using
the transformation 
\begin{equation}
U_{{\rm rot}}=e^{-i\left(\phi+\pi/2\right)\sigma^{z}/2}.\label{eq:Urot}
\end{equation}
Letting $a\equiv i\tilde{a},$ we can express the rotated Hamiltonian
as 
\begin{eqnarray}
H_{r}^{{\rm rot}} & = & U_{{\rm rot}}^{\dagger}H_{r}^{{\rm rf}}U_{{\rm rot}}\nonumber \\
 & = & \Delta_{0}a^{\dagger}a+g\left(e^{i\phi}\sigma^{+}a+e^{-i\phi}\sigma^{-}a^{\dagger}\right)+\Omega\sigma^{y}.\label{eq:Hrrot}
\end{eqnarray}

Eq. (\ref{eq:Hrrot}) gives the Hamiltonian for one RX qubit coupled
to a resonator and driven by an external microwave field, in a frame
rotating at its resonance frequency $\nu=\omega.$ In order to obtain
interaction terms that generate sideband transitions, we now transform
$H_{r}^{{\rm rot}}$ to a second frame rotating at the Rabi frequency
$2\Omega$ for the qubit and at the effective frequency $\Delta_{0}$
for the resonator using
\begin{equation}
U_{2}=e^{-i\left(\Delta_{0}a^{\dagger}a+\Omega\sigma^{y}\right)\tau}.\label{eq:U2}
\end{equation}
This unitary transformation yields
\begin{eqnarray}
H_{r}^{{\rm drf}} & = & \frac{g}{2}\left[\cos\left(2\Omega\tau\right)\sigma^{x}+\sin\left(2\Omega\tau\right)\sigma^{z}\right]\nonumber \\
 & \times & \left(e^{i\phi}e^{-i\Delta_{0}\tau}a+e^{-i\phi}e^{i\Delta_{0}\tau}a^{\dagger}\right)\nonumber \\
 & + & i\frac{g}{2}\sigma^{y}\left(e^{i\phi}e^{-i\Delta_{0}\tau}a-e^{-i\phi}e^{i\Delta_{0}\tau}a^{\dagger}\right),\label{eq:Hrdrf}
\end{eqnarray}
which describes the interaction of the RX qubit with the resonator
in terms of a time-dependent rotation of the transition dipole moment
induced by the Rabi frequency. Finally, applying the transformation
\begin{equation}
U_{{\rm rot}}^{\prime}=e^{i\left(\pi/4\right)\sigma^{x}},\label{eq:Urotprime}
\end{equation}
to $H_{r}^{{\rm drf}}$ yields
\begin{eqnarray}
\tilde{H}_{r}^{{\rm drf}} & = & \frac{g}{2}\left[e^{i\phi}e^{-i\left(\Delta_{0}-2\Omega\right)\tau}\sigma^{+}a\ +\ e^{-i\phi}e^{i\left(\Delta_{0}-2\Omega\right)\tau}\sigma^{-}a^{\dagger}\right.\nonumber \\
 & + & \left.e^{-i\phi}e^{i\left(\Delta_{0}+2\Omega\right)\tau}\sigma^{+}a^{\dagger}\ +\ e^{i\phi}e^{-i\left(\Delta_{0}+2\Omega\right)\tau}\sigma^{-}a\right]\nonumber \\
 & + & i\frac{g}{2}\sigma^{z}\left(e^{i\phi}e^{-i\Delta_{0}\tau}a-e^{-i\phi}e^{i\Delta_{0}\tau}a^{\dagger}\right).\label{eq:Hrdrftilde}
\end{eqnarray}

\begin{figure}
\includegraphics[bb=0bp 10bp 335bp 200bp,width=2.5in]{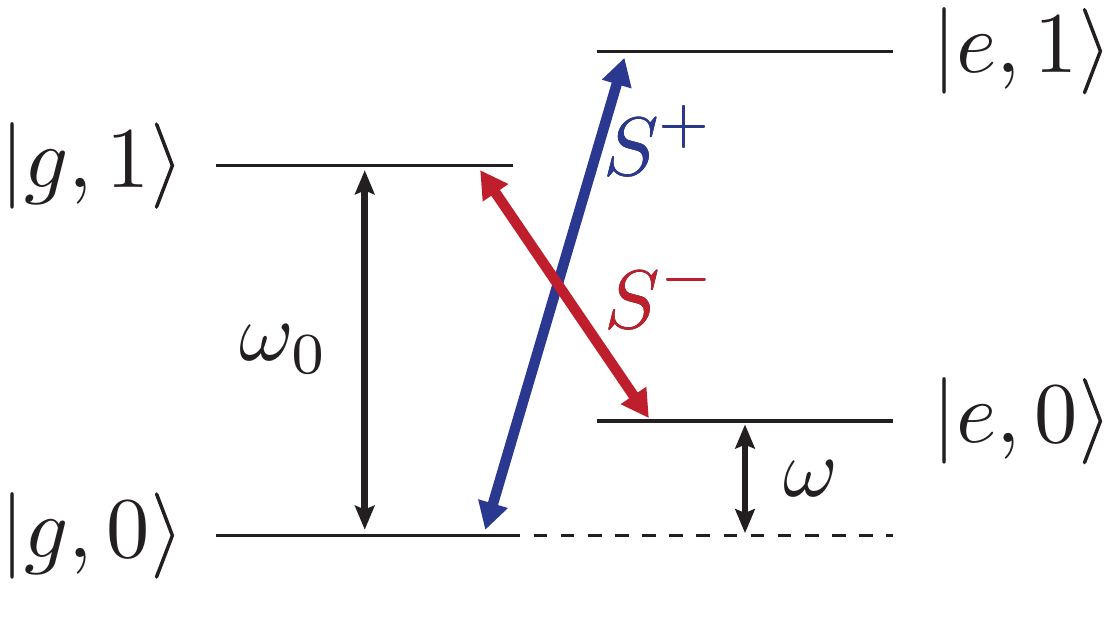}

\protect\caption{\label{fig:sidebands}Red ($S^{-}$) and blue ($S^{+}$) sideband
transitions {[}Eqs. (\ref{eq:Sminus}) and (\ref{eq:Splus}), respectively{]}
generated in the doubly rotating frame for a single RX qubit coupled
to a resonator and driven on resonance ($\nu=\omega$).}
\end{figure}

Based on the form of Eq. (\ref{eq:Hrdrftilde}), we make a rotating
wave approximation for two cases. First, we set $\Delta_{0}=2\Omega$
in Eq. (\ref{eq:Hrdrftilde}) and neglect the rapidly oscillating
terms containing $\sigma^{+}a^{\dagger},$ $\sigma^{-}a,$ $\sigma^{z}a,$
and $\sigma^{z}a^{\dagger}$ to obtain 
\begin{equation}
H_{-}=\frac{g}{2}\left(e^{i\phi}\sigma^{+}a+e^{-i\phi}\sigma^{-}a^{\dagger}\right).\label{eq:Hminus}
\end{equation}
Alternatively, setting $\Delta_{0}=-2\Omega,$ we find that the rapidly
oscillating terms are those containing $\sigma^{+}a,$ $\sigma^{-}a^{\dagger},$
$\sigma^{z}a,$ and $\sigma^{z}a^{\dagger}.$ After dropping these
terms, Eq. (\ref{eq:Hrdrftilde}) becomes
\begin{equation}
H_{+}=\frac{g}{2}\left(e^{-i\phi}\sigma^{+}a^{\dagger}+e^{i\phi}\sigma^{-}a\right).\label{eq:Hplus}
\end{equation}
The Hamiltonians $H_{-}$ and $H_{+}$ generate ``red'' and ``blue''
sideband transitions, respectively \cite{Childs2000,Blais2007}, as
can be verified by considering their action on the qubit-resonator
basis states $\ket{g,n}$ and $\ket{e,n}$ (Fig. \ref{fig:sidebands}).
The gates generated by $H_{\pm}$ can be expressed as \cite{Childs2000,Blais2007}
\begin{eqnarray}
S^{-}\left(\theta,\phi\right) & = & e^{-i\left(\theta/2\right)\left(e^{i\phi}\sigma^{+}a+e^{-i\phi}\sigma^{-}a^{\dagger}\right)},\label{eq:Sminus}\\
S^{+}\left(\theta,\phi\right) & = & e^{-i\left(\theta/2\right)\left(e^{-i\phi}\sigma^{+}a^{\dagger}+e^{i\phi}\sigma^{-}a\right)},\label{eq:Splus}
\end{eqnarray}
where $\theta\equiv g\tau.$ Thus, although $\omega\neq\omega_{0},$
driving each RX qubit resonantly and moving to a doubly rotating frame
via $U_{1}$ and $U_{2}$ {[}Eqs. (\ref{eq:U1}) and (\ref{eq:U2}){]}
generates sideband transitions that transfer excitations between the
qubit and the resonator via an effective resonance in the doubly rotating
frame that is enabled by the combination of the drive and resonator
photons. 

We now construct a controlled-phase gate for two RX qubits using Eqs.
(\ref{eq:Sminus}) and (\ref{eq:Splus}). Here, we focus on the controlled-Z
($\pi$-phase) gate. Following the approach of Ref. \citenum{Childs2000},
we use a sequence of red sideband transitions to define the gate 
\begin{equation}
W\equiv S^{-}\left(\frac{\pi}{2},0\right)S^{-}\left(\pi\sqrt{2},\frac{\pi}{2}\right)S^{-}\left(-\frac{\pi}{2},0\right),\label{eq:W}
\end{equation}
which acts on a single qubit coupled to the resonator. As discussed
in Ref. \citenum{Childs2000}, this gate plays a role equivalent to
the $2\pi$ pulse used in the Cirac-Zoller sequence for trapped ions
\cite{Cirac1995} but requires only the two levels of a qubit, in
contrast to resonant approaches proposed for superconducting qubits
\cite{Haack2010}. As $W$ also prevents leakage to resonator states
other than $\ket{0}$ and $\ket{1}$, we can consider the subspace
$\left\{ \ket{g,0},\ket{g,1},\ket{e,0},\ket{e,1}\right\} .$ In this
basis, Eq. (\ref{eq:W}) has the form 
\begin{equation}
W=\left(\begin{array}{cccc}
1\\
 & e^{-i\pi/\sqrt{2}}\\
 &  & e^{i\pi/\sqrt{2}}\\
 &  &  & -1
\end{array}\right).\label{eq:Wmat}
\end{equation}
Combining the gate $W$ with additional blue sideband transitions
and single-qubit $z$-axis rotations yields the gate sequence
\begin{equation}
U_{{\rm CZ}}=R_{z,a}\left(\frac{\pi}{\sqrt{2}}\right)S_{a}^{+}\left(\pi,\phi\right)W_{b}R_{z,b}\left(\frac{\pi}{\sqrt{2}}\right)S_{a}^{+}\left(\pi,\phi\right),\label{eq:Ucz}
\end{equation}
where the added subscripts $a,b$ indicate the qubit on which each
gate acts. In the zero-photon subspace $\left\{ \ket{e,e,0},\ket{e,g,0},\ket{g,e,0},\ket{g,g,0}\right\} $,
$U_{{\rm CZ}}$ takes the form of a controlled-Z gate with respect
to the two-qubit basis states and thus entangles the two qubits. 

Note that, in addition to $W$ and additional sideband transitions,
the full sequence for the controlled-Z gate in Eq. (\ref{eq:Ucz})
requires single-qubit rotations about the $z$ axis of the Bloch sphere
for each RX qubit. As seen from Eq. (\ref{eq:Hrx}), rapid single-qubit
rotations for the RX qubit can be generated via exchange \cite{Medford2013,Taylor2013}.
In the absence of detuning variations about the operating point (i.e.,
for $F=0$), $H_{RX}$ generates a rotation $R_{z}\left(\varphi\right)\equiv e^{-i(\varphi/2)\sigma^{z}}$
about the $z$ axis of the Bloch sphere of a RX qubit, with $\varphi=\omega\tau.$
With qubit-resonator coupling present, a $z$-axis rotation for a
particular qubit can be carried out by tuning $\omega$ such that
$\left|\tilde{\Delta}\right|\equiv\left|\omega_{0}-\omega\right|\gg g,$
so that $g/\left|\tilde{\Delta}\right|\rightarrow0$ and the interaction
of the qubit with the resonator effectively vanishes. 

While the qubit frequency must be well-separated from the resonator
frequency $\omega_{0},$ we note that $\omega\gg g$ is also satisfied
for typical system parameters (e.g., $\omega\sim2\pi\times1\ {\rm GHz}$
\cite{Taylor2013} and $g\lesssim2\pi\times10\ {\rm MHz}$). Thus,
we expect the two-qubit gate rate to be limited by that of the sideband
gates. As all sideband gates in Eq. (\ref{eq:Ucz}) are generated
by interactions that depend linearly on the coupling strength $g$
via $\theta=g\tau,$ and as the exchange-generated single-qubit rotations
have a rate $\omega\gg g,$ this suggests that the overall two-qubit
gate rate scales as $g\sim\xi^{2}.$ The driven resonant regime should
therefore enable more rapid gates than those discussed in Sec. \ref{sub:disp}.
We note that the resonant regime of quantum dot-cavity coupling has
been achieved in recent experiments \cite{Kim2014PRL}, which demonstrate
measurements of the transition to this regime via the observation
of an enhanced intensity of the resonant sideband of the Mollow triplet
characteristic of a strongly driven two-level system \cite{Mollow1969}.

\subsection{\label{sub:drivdisp}Driven dispersive regime }

Generating entanglement between RX qubits in the presence of variation
in their transition frequencies enables both two-qubit gates and addressability
of individual qubits. Thus, we now consider the dispersive regime
for two qubits coupled to the resonator with $\omega_{a}\neq\omega_{b}.$
In contrast to the approach of Sec. \ref{sub:disp}, we also include
a microwave driving field acting on the resonator (see Sec. \ref{sub:resonant}
and Fig. \ref{fig:RXresRX}) with a frequency $\nu$ equal to the
transition frequency of one qubit. Interaction with the resonator
shifts both the Rabi frequency of the qubit resonant with the driving
field and the difference of the qubit transition frequencies. We derive
an entangling gate that is generated by the effective interaction
in a frame rotating at both the drive frequency and the modified qubit
Rabi and difference frequencies. This doubly rotating frame enables
energy exchange between the qubits even when they have different transition
frequencies, in analogy to Hartmann-Hahn double resonance in NMR \cite{Hartmann1962,Rigetti2005}.

From Eqs. (\ref{eq:H0}) and (\ref{eq:V}), the Hamiltonian is given
by 
\begin{eqnarray}
H_{dd} & = & \omega_{0}a^{\dagger}a+\sum_{\mu=a,b}\frac{\omega_{\mu}}{2}\sigma_{\mu}^{z}\nonumber \\
 & + & \sum_{\mu=a,b}g_{\mu}\sigma_{\mu}^{x}\left(a+a^{\dagger}\right)\nonumber \\
 & + & \varepsilon\left[e^{-i\left(\nu\tau+\phi\right)}a^{\dagger}+e^{i\left(\nu\tau+\phi\right)}a\right],\label{eq:Hdd}
\end{eqnarray}
We displace the resonator field using $D\left(\alpha\right)$ as in
Eq. (\ref{eq:Hrprime}) in order to eliminate the direct action of
the driving field on the resonator, obtaining

\begin{eqnarray}
H_{dd}^{\prime} & = & \omega_{0}a^{\dagger}a+\sum_{\mu=a,b}\frac{\omega_{\mu}}{2}\sigma_{\mu}^{z}\nonumber \\
 & + & \sum_{\mu=a,b}g_{\mu}\sigma_{\mu}^{x}\left(a+a^{\dagger}\right)\nonumber \\
 & - & \sum_{\mu=a,b}2\Omega_{\mu}\cos\left(\nu t+\phi\right)\sigma_{\mu}^{x},\label{eq:Hddprime}
\end{eqnarray}
where $2\Omega_{\mu}\equiv2g_{\mu}\varepsilon/\Delta_{0}.$ 

We next transform to a rotating frame via 
\begin{eqnarray}
U_{1}^{\prime} & = & e^{-i\nu\left(a^{\dagger}a+\sigma_{a}^{z}/2+\sigma_{b}^{z}/2\right)\tau}.\label{eq:U1prime}
\end{eqnarray}
Subsequently assuming the driving field is resonant with the transition
of qubit $a$ such that $\nu=\omega_{a},$ setting $\Omega_{b}=\phi=0$
for simplicity, and dropping rapidly oscillating terms $\sim e^{\pm2i\nu\tau}$
leads to
\begin{eqnarray}
H_{dd}^{{\rm rf}} & \approx & \Delta_{0}a^{\dagger}a-\Omega_{a}\sigma_{a}^{x}+\frac{\delta}{2}\sigma_{b}^{z}\nonumber \\
 & + & \sum_{\mu=a,b}g_{\mu}\left(\sigma_{\mu}^{+}a+\sigma_{\mu}^{-}a^{\dagger}\right),\label{eq:Hddrf}
\end{eqnarray}
where we have defined the qubit frequency difference $\delta\equiv\omega_{b}-\omega_{a}=\omega_{b}-\nu.$
We also apply a rotation to qubit $a$ using 
\begin{equation}
U_{a}=e^{i\left(\pi/4\right)\sigma_{a}^{y}}\label{eq:Ua}
\end{equation}
in order to diagonalize the term with $\sigma_{a}^{x}$ in the second
line of Eq. (\ref{eq:Hddrf}). This transformation yields a Hamiltonian
$H_{dd}^{{\rm rot}}=U_{a}^{\dagger}H_{dd}^{{\rm rf}}U_{a}$ which
contains additional terms not present in the Hamiltonian $H_{d}$
considered in Sec. \ref{sub:disp} for two qubits of equal transition
frequencies in the dispersive regime and in the absence of a driving
field. In order to simplify the analysis leading to an effective interaction
between the two RX qubits in the present case, we apply perturbation
theory and take into account only states within the low-energy subspace
$\left\{ \ket{0},\ket{1}\right\} $ for the resonator. We write $H_{dd}^{{\rm rot}}=H_{0}^{\prime}+V_{dd},$
where
\begin{eqnarray}
H_{0}^{\prime} & = & \Delta_{0}a^{\dagger}a+\Omega_{a}\sigma_{a}^{z}+\frac{\delta}{2}\sigma_{b}^{z},\label{eq:H0prime}\\
V_{dd} & = & -\frac{g_{a}}{2}\sigma_{a}^{z}\left(a+a^{\dagger}\right)\nonumber \\
 &  & +\frac{g_{a}}{2}\left(\sigma_{a}^{+}-\sigma_{a}^{-}\right)\left(a-a^{\dagger}\right)\nonumber \\
 &  & +g_{b}\left(\sigma_{b}^{+}a+\sigma_{b}^{-}a^{\dagger}\right).\label{eq:Vdd}
\end{eqnarray}
Assuming $g_{a,b}\ll\Omega_{a}\approx\left|\delta\right|\ll\left|\Delta_{0}\right|,$
we consider separately the three subspaces defined by the projection
operators $P_{+}\equiv\ket{e,e,0}\bra{e,e,0},$ $P_{0}\equiv\ket{e,g,0}\bra{e,g,0}+\ket{g,e,0}\bra{g,e,0},$
and $P_{-}\equiv\ket{g,g,0}\bra{g,g,0}.$ Denoting the the projector
for the one-photon subspace by $Q,$ the resulting effective Hamiltonian
in the full zero-photon subspace $P\equiv P_{+}+P_{0}+P_{-}$ is found
to be 
\begin{eqnarray}
H_{{\rm eff}}^{\left(0\right)} & \equiv & PH_{{\rm eff}}P\nonumber \\
 & = & PH_{0}^{\prime}P+PV_{dd}Q\frac{1}{\varepsilon{}_{0}-QH_{0}^{\prime}Q}QV_{dd}P\nonumber \\
 & = & \tilde{\Omega}_{a}\sigma_{a}^{z}+\frac{\tilde{\delta}}{2}\sigma_{b}^{z}\nonumber \\
 & - & \frac{g_{a}g_{b}}{2\left(\Delta_{0}-\delta\right)}\left(\sigma_{a}^{+}\sigma_{b}^{-}+\sigma_{a}^{-}\sigma_{b}^{+}\right),\nonumber \\
\label{eq:Heff0}
\end{eqnarray}
where we have set $2\Omega_{a}=\delta$ for simplicity, chosen $\varepsilon_{0}=0,\pm\delta$
for $P_{0,\pm},$ respectively, and defined the quantities 
\begin{eqnarray}
2\tilde{\Omega}_{a} & \equiv & \delta-\frac{g_{a}^{2}}{4}\left(\frac{1}{\Delta_{0}-\delta}-\frac{1}{\Delta_{0}+\delta}\right),\label{eq:2Omegaatilde}\\
\tilde{\delta} & \equiv & \delta-\frac{g_{b}^{2}}{\Delta_{0}-\delta}.\label{eq:deltaatilde}
\end{eqnarray}
Here, $2\tilde{\Omega}_{a}$ represents the effective Rabi frequency
of RX qubit $a$ and $\tilde{\delta}$ is the effective qubit transition
frequency difference when both RX qubits are coupled to the resonator
in the dispersive regime. 

Finally, we transform to a second rotating frame using the operator
\begin{eqnarray}
U_{2}^{\prime} & = & e^{-i\left(\tilde{\Omega}_{a}\sigma_{a}^{z}+\tilde{\delta}\sigma_{b}^{z}/2\right)\tau},\label{eq:U2prime}
\end{eqnarray}
 which acts within the zero-photon subspace. This leads to
\begin{eqnarray}
\tilde{H}_{{\rm eff}}^{\left(0\right)} & = & -\frac{g_{a}g_{b}}{2\left(\Delta_{0}-\delta\right)}\nonumber \\
 & \times & \left[\sigma_{a}^{+}\sigma_{b}^{-}e^{i\left(2\tilde{\Omega}_{a}-\tilde{\delta}\right)\tau}+\sigma_{a}^{-}\sigma_{b}^{+}e^{-i\left(2\tilde{\Omega}_{a}-\tilde{\delta}\right)\tau}\right]\label{eq:Heff0tilde}
\end{eqnarray}
We now set $2\tilde{\Omega}_{a}=\tilde{\delta},$ which from Eqs.
(\ref{eq:2Omegaatilde}) and (\ref{eq:deltaatilde}) also leads to
the constraint 
\begin{equation}
g_{b}=g_{a}\sqrt{\frac{\delta}{2\left(\Delta_{0}+\delta\right)}}\label{eq:gconstraint}
\end{equation}
relating the qubit-resonator coupling strengths. For the effective
Hamiltonian in the doubly rotating frame, we then find 
\begin{equation}
H_{dd}^{{\rm drf}}\approx-\frac{g_{a}g_{b}}{2\left(\Delta_{0}-\delta\right)}\left(\sigma_{a}^{+}\sigma_{b}^{-}+\sigma_{a}^{-}\sigma_{b}^{+}\right),\label{eq:Hdddrf}
\end{equation}
where $g_{b}$ and $g_{a}$ are related by Eq. (\ref{eq:gconstraint})
for the case $2\Omega_{a}=\delta.$ Thus, driving qubit $a$ such
that its Rabi frequency is resonant with the difference between the
qubit transition frequencies leads to an effective two-qubit interaction
in the doubly rotating frame. Note that this interaction has the same
form as (\ref{eq:Hdrfapprox}). The unitary evolution generated by
$H_{dd}^{{\rm drf}}$ is therefore of the same form as Eq. (\ref{eq:Hdrfapprox})
and gives rise to the same two-qubit entangling gates, with a rate
$g_{a}g_{b}/2\left(\Delta_{0}-\delta\right)\sim\sqrt{\delta/\Delta_{0}}\left(g^{2}/\Delta_{0}\right)\sim\xi^{4},$
while additionally allowing for variation in the transition frequencies
of the two qubits.

\section{\label{sec:ImplementSi}Implementation in Si triple quantum dots}

We now consider the feasibility of implementing the approaches for
entangling RX qubits discussed in the present work within silicon
quantum dots. Recent work has demonstrated rapid, coherent control
of a single RX qubit in a GaAs triple dot \cite{Medford2013}. However,
in addition to gate voltage noise, the coherence is expected to be
further limited in GaAs by the nuclear spin environment and by piezoelectric
phonons \cite{Taylor2013}, which typically represent the dominant
sources of charge and spin relaxation in GaAs quantum dots \cite{Hanson2007RMP}.
Implementing RX qubits in Si potentially provides improved coherence
due to a low nuclear spin concentration, which can be made to approach
zero via isotopic purification, as well as to the absence of piezoelectric
phonons \cite{Zwanenburg2013,Kawakami2014,Muhonen2014,Eng2015,Veldhorst2015,Takeda2016arxiv,Kawakami2016arxiv}.
However, the valley degree of freedom in Si leads to a more complex
low-energy spectrum than that of GaAs. Thus, a direct extension of
RX qubit properties to Si is not obviously straightforward. 

Here, we focus on the effects of relaxation due to electron-phonon
coupling in the context of the approaches described in the present
work. We assume that each RX qubit is coupled independently to the
same phonon bath and calculate the relaxation rate for a single three-electron
triple quantum dot in Si. In order to identify the most relevant relaxation
transition, we note that a typical single-dot valley splitting $E_{{\rm V}}\gtrsim100\ \mu{\rm eV}$
\cite{Zwanenburg2013,Yang2013}, while typical qubit frequencies that
set the gap between the qubit basis states $\ket{g}$ and $\ket{e}$
correspond to $\hbar\omega<10\ \mu{\rm eV}.$ Thus, we assume that
$\omega$ determines the lowest relevant gap for the RX qubit and
that the dominant relaxation process is charge relaxation from $\ket{e}$
to $\ket{g}$ within the lowest-energy valley manifold. The calculation
is then similar to that performed in Ref. \citenum{Taylor2013} for
the case of a GaAs triple dot. 

Unlike GaAs, however, the crystal structure of unstrained Si has a
center of inversion symmetry, which leads to the absence of piezoelectric
phonons. The electron-phonon interaction for Si therefore consists
only of deformation potential terms. In the presence of strain along
the {[}001{]} ($z^{\prime}$) axis, we can write this interaction
as \cite{Yu2010} 
\begin{eqnarray}
H_{{\rm ep}} & = & \Xi_{d}{\bf \nabla}\cdot{\bf u}+\Xi_{u}\frac{\partial u_{z^{\prime}}}{\partial z^{\prime}},\label{eq:Hep}
\end{eqnarray}
where $\Xi_{d}$ and $\Xi_{u}$ are the dilation and uniaxial deformation
potentials, respectively, and the phonon displacement vector ${\bf u}$
is given by 
\begin{eqnarray}
{\bf u}\left({\bf r}\right) & = & \sum_{\mu,{\bf k}}\sqrt{\frac{\hbar}{2\rho_{0}V_{0}c_{\mu}k}}\left(a_{\mu,{\bf k}}+a_{\mu,-{\bf k}}^{\dagger}\right)e^{i{\bf k}\cdot{\bf r}}\ \hat{\boldsymbol{\epsilon}}_{\mu,{\bf k}}.\label{eq:u}
\end{eqnarray}
Here, the operator $a_{\mu,\mathbf{k}}^{\dagger}$ ($a_{\mu,\mathbf{k}}$)
creates (annihilates) an acoustic phonon with wave vector $\mathbf{k}$,
polarization $\mu$ {[}the sum in Eq. (\ref{eq:u}) is taken over
one longitudinal mode ($\mu=l$) and two transverse modes ($\mu=p$){]},
phonon speed $c_{\mu},$ energy $\varepsilon_{\mathrm{ph}}=\hbar c_{\mu}k,$
and unit polarization vector $\hat{\boldsymbol{\epsilon}}_{\mu,{\bf k}},$
$\rho_{0}$ is the mass density of the material, and $V_{0}$ is the
crystal volume. Evaluating the derivatives in $H_{{\rm ep}}$ leads
to 
\begin{eqnarray}
H_{{\rm ep}} & = & i\sum_{\mu,{\bf k}}\sqrt{\frac{\hbar}{2\rho_{0}V_{0}c_{\mu}k}}\left({\bf k}\cdot\hat{\boldsymbol{\epsilon}}_{\mu,{\bf k}}\ \Xi_{d}\right.\label{eq:Hepexpr}\\
 & + & \left.\ k_{z^{\prime}}\,\hat{z}^{\prime}\cdot\hat{\boldsymbol{\epsilon}}_{\mu,{\bf k}}\ \Xi_{u}\right)\left(a_{\mu,{\bf k}}+a_{\mu,-{\bf k}}^{\dagger}\right)M_{k},
\end{eqnarray}
where the factor 
\begin{eqnarray}
M_{k} & \equiv & \sum_{i,j=1}^{3}\sum_{\sigma}\bra{i}e^{i\mathbf{k}\cdot\mathbf{r}}\ket{j}c_{i\sigma}^{\text{\dag}}c_{j\sigma}\label{eq:Mk}
\end{eqnarray}
contains the dependence on electronic degrees of freedom and ${\bf r}$
is the electron position operator.

The rate $\Gamma$ of qubit relaxation due to $H_{\mathrm{ep}}$ is
given by Fermi's golden rule as $\Gamma\sim\left|\left\langle g\right|H_{\mathrm{ep}}\left|e\right\rangle \right|^{2}\rho\left(\omega\right)$,
where $\rho\left(\omega\right)$ is the phonon density of states evaluated
at the exchange gap $\omega$ between the logical qubit states $\left|g\right\rangle $
and $\left|e\right\rangle $ that determines the energy of the emitted
phonon. We can write the relaxation rate as $\Gamma=s_{l}I_{l}\left(\omega/\hbar c_{l}\right)+s_{p}I_{p}\left(\omega/\hbar c_{p}\right),$
which is expressed in terms of the momentum-space angular integrals
\begin{eqnarray}
I_{l}\left(k\right) & \equiv & \int\left(1+\Lambda\cos^{2}\beta\right)^{2}\left|\bra{g}M_{k}\ket{e}\right|^{2}d\Omega_{{\rm ang}},\label{eq:Ilk}\\
I_{p}\left(k\right) & \equiv & \int\Lambda^{2}\cos^{2}\beta\sin^{2}\beta\left|\bra{g}M_{k}\ket{e}\right|^{2}d\Omega_{{\rm ang}}\label{eq:Ipk}
\end{eqnarray}
and the factors 
\begin{eqnarray}
s_{\mu} & = & \frac{\omega^{3}}{8\pi^{2}\hbar^{4}\rho_{0}c_{\mu}^{5}}\Xi_{d}^{2},\ \ \mu=l,p.\label{eq:smu}
\end{eqnarray}
In order to obtain Eqs. (\ref{eq:Ilk}) and (\ref{eq:Ipk}), we have
chosen one of the two transverse ($\mu=p$) polarization axes to lie
perpendicular to the direction of uniaxial strain $\hat{z}^{\prime}$
and defined $\beta$ as the angle between ${\bf k}$ and $\hat{z}^{\prime}.$
We have also defined the dimensionless parameter $\Lambda\equiv\Xi_{u}/\Xi_{d}$
and used $\Omega_{{\rm ang}}$ to denote the momentum-space solid
angle. 

\begin{figure}
\includegraphics[bb=0bp 50bp 308bp 283bp,width=3.2in]{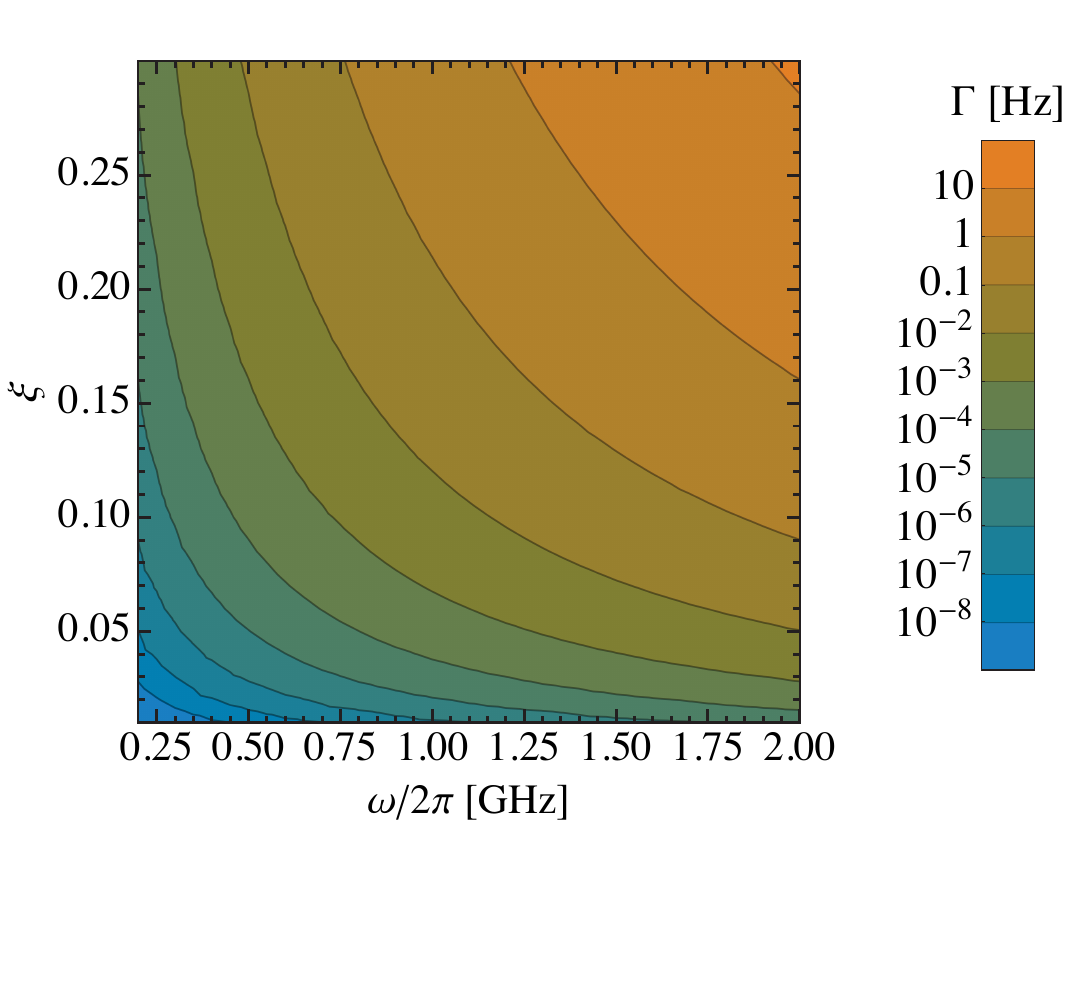}

\protect\caption{\label{fig:relaxationrate}Contour plot of the phonon-induced relaxation
rate $\Gamma$ at the operation point $\epsilon_{0}=0,$ as a function
of the qubit frequency $\omega$ and the charge admixture parameter
$\xi=t/\Delta.$ }
\end{figure}

The relaxation rate at the RX qubit operation point $\epsilon_{0}=0$
is plotted in Fig. \ref{fig:relaxationrate} as a function of the
exchange gap $\omega$ and the charge admixture parameter $\xi$.
The parameters used to calculate $\Gamma$ are the Si transverse effective
mass $m^{*}=0.19\,m_{e},$ the single-dot size $\sigma=23\ \mbox{nm}$
and left-right dot separation $w=260\ \mbox{nm}$ \cite{Taylor2013},
$\rho_{0}=2.33\times10^{3}\ \mbox{kg/m}^{3},$ $c_{l}=9.33\times10^{3}\ \mathrm{m}/\mathrm{s},$
$c_{p}=5.42\times10^{3}\ \mathrm{m}/\mathrm{s},$ $\Xi_{d}=5\ \mbox{eV},$
and $\Xi_{u}=8.77\ \mbox{eV}$ \cite{Tahan2014}. 

Comparing Fig. \ref{fig:relaxationrate} with Fig. 2(e) in Ref. \citenum{Taylor2013}
reveals that the rates for phonon-induced relaxation of the resonant
exchange qubit in Si are smaller than those for GaAs by several orders
of magnitude. For the ranges of $\omega$ and $\xi$ shown in the
figure, the relaxation rate is well approximated by $\Gamma=\frac{\omega^{5}}{\nu_{0}^{4}}\xi^{4}$
with the fit parameter $\nu_{0}=2\pi\times110\ {\rm GHz.}$ We thus
find that the relaxation rate for Si exhibits a $\sim\omega^{5}$
exchange gap dependence, in contrast to the $\sim\omega^{3}$ dependence
for GaAs (see Ref. \citenum{Taylor2013}). We note in particular that
$\Gamma\lesssim10\ {\rm Hz}$ for all values of $\xi$ and $\omega$
shown, which is much smaller than typical values of $T_{2}^{\ast-1}.$
Thus, in contrast to GaAs implementations \cite{Taylor2013}, the
coherence time for a RX qubit in a Si triple quantum dot is unlikely
to be limited by phonon-induced decay. Si-based RX qubits should therefore
enable the strong coupling regime to be achieved for larger $\xi^{2}$,
which in principle leads to more rapid and robust entangling gates.

\section{\label{sec:EntanglingF}Performance of entangling gates}

Finally, we consider the performance of the two-qubit entangling gates
for RX qubits discussed in Sec. \ref{sec:RXresRX} and calculate gate
fidelities for each of the three regimes. As we have seen for RX qubits
in Si triple quantum dots, the absence of piezoelectric phonons is
expected to lead to qubit relaxation times which are much longer than
qubit dephasing times (see Sec. \ref{sec:ImplementSi} and Fig. \ref{fig:relaxationrate}).
We therefore assume in the present analysis that the dominant decay
processes are pure dephasing of the qubits with rates $\gamma_{a,b}$
and the decay of photons out of the resonant cavity with rate $\kappa.$
Throughout this section, we focus on the strong coupling regime $\gamma_{a,b},\kappa<g$
of the qubit-resonator interactions and assume that $\omega_{0}>\omega_{\mu}$
for $\mu=a,b.$

\subsection{\label{sub:Fdisp}Dispersive regime}

We first consider the dispersive regime, defined by $g_{\mu}\ll\tilde{\Delta}_{\mu}=\omega_{0}-\omega_{\mu}$
(see Sec. \ref{sub:disp}). After making a rotating wave approximation
for $\tilde{\Delta}_{\mu}\ll\omega_{0}+\omega_{\mu},$ the Hamiltonian
describing the system is $H_{d}=H_{0}+V,$ where $H_{0}$ and $V$
are given by Eqs. (\ref{eq:H0}) and (\ref{eq:Vrwa}), respectively.
We describe the corresponding time evolution in the presence of qubit
dephasing and cavity decay by the master equation \cite{Blais2007,Lambropoulos2006}
\begin{eqnarray}
\dot{\rho_{d}} & = & -i\left[H_{d},\rho_{d}\right]+\sum_{\mu=a,b}\frac{\gamma_{\mu}}{2}\left(\sigma_{\mu}^{z}\rho_{d}\sigma_{\mu}^{z}-\rho_{d}\right)\nonumber \\
 & + & \frac{\kappa}{2}\left(2a\rho_{d}a^{\dagger}-a^{\dagger}a\rho_{d}-\rho_{d}a^{\dagger}a\right),\label{eq:mastereqn}
\end{eqnarray}
where $\rho_{d}$ represents the density matrix of the combined system
consisting of both qubits and the resonator. In order to obtain the
gate fidelity, we compare the solution of Eq. (\ref{eq:mastereqn})
with that for the ideal evolution (for which $\gamma_{a,b}=\kappa=0$). 

For simplicity, we set $\omega_{a}=\omega_{b}\equiv\omega$ (which
corresponds to $\tilde{\Delta}_{a}=\tilde{\Delta}_{b}\equiv\tilde{\Delta}$),
$g_{a}=g_{b}\equiv g,$ and $\gamma_{a}=\gamma_{b}\equiv\gamma$ in
what follows. As we consider the dispersive regime of qubit-resonator
coupling, we also confine our description of the resonator mode to
the $n=0$ and $n=1$ photon subspaces. We can then write 
\begin{eqnarray}
h_{0} & \equiv & PH_{d}P=PH_{0}P=\frac{\omega}{2}\left(\sigma_{a}^{z}+\sigma_{b}^{z}\right),\label{eq:h0}\\
h_{1} & \equiv & QH_{d}Q=QH_{0}Q=\omega_{0}+\frac{\omega}{2}\left(\sigma_{a}^{z}+\sigma_{b}^{z}\right),\label{eq:h1}\\
v & \equiv & PH_{d}Q=PVQ=g\left(\sigma_{a}^{+}+\sigma_{b}^{+}\right),\label{eq:v}
\end{eqnarray}
where $P$ and $Q$ are the zero-photon and one-photon projectors
defined in Sec. (\ref{sub:drivdisp}). Using these quantities, we
can re-express the master equation {[}Eq. (\ref{eq:mastereqn}){]}
in terms of the subspace projections $\rho_{00}\equiv P\rho P,$ $\rho_{11}\equiv Q\rho Q,$
$\rho_{01}\equiv P\rho Q,$ and $\rho_{10}=\rho_{01}^{\dagger}=Q\rho P$
of the density matrix. 

To simplify the analysis, we set $\rho_{11}=0,$ which amounts to
neglecting the contribution of the quantum jump term $\kappa a\rho a^{\dagger}$
in Eq. (\ref{eq:mastereqn}) (this approximation is reasonable for
$\kappa,g\ll\tilde{\Delta}$). We also assume that the photon coherences
decay much more rapidly than the photon populations and neglect the
time evolution of the photon coherences by setting $\dot{\rho}_{01}=0.$
Note that in the absence of decay, setting this condition is equivalent
to carrying out perturbation theory for the zero-photon subspace.
From Eq. (\ref{eq:mastereqn}), we then find 
\begin{eqnarray}
\dot{\rho}_{00} & = & -i\left(\left[h_{0},\rho_{00}\right]+v\rho_{01}^{\dagger}-\rho_{01}v^{\dagger}\right)\nonumber \\
 & + & \frac{\gamma}{2}\sum_{\mu=a,b}\left(\sigma_{\mu}^{z}\rho_{00}\sigma_{\mu}^{z}-\rho_{00}\right),\label{eq:rho00meq}\\
\dot{\rho}_{01} & = & -i\left(h_{0}\rho_{01}-\rho_{01}h_{1}-\rho_{00}v\right)\nonumber \\
 & + & \frac{\gamma}{2}\sum_{\mu=a,b}\left(\sigma_{\mu}^{z}\rho_{01}\sigma_{\mu}^{z}-\rho_{01}\right)-\frac{\kappa}{2}\rho_{01}.\label{eq:rho01meq}
\end{eqnarray}
Performing a mapping to a Liouville-space representation {[}i.e.,
expressing the density matrix projections as vectors and the superoperator
terms in Eqs. (\ref{eq:rho00meq}) and (\ref{eq:rho01meq}) as matrices{]}
and setting $\dot{\rho}_{01}=0$ enables $\rho_{01}$ to be expressed
in terms of $\rho_{00}.$ Solving Eq. (\ref{eq:rho00meq}) with this
relation substituted for $\rho_{01}$ then yields an analytical solution
for the density matrix $\rho_{00}\left(\tau\right)$ as a function
of time $\tau.$ 

In order to calculate the gate fidelity, we choose as the initial
state of the system 
\begin{equation}
\rho_{00}\left(0\right)=\ket{e,g,0}\bra{e,g,0}.\label{eq:rho0}
\end{equation}
For the ideal evolution, described by setting $\gamma=\kappa=0$ in
Eq. (\ref{eq:mastereqn}), we find 
\begin{equation}
\rho_{00,{\rm id}}\left(\tau\right)\equiv U_{d}\left(\tau\right)\rho_{00}\left(0\right)U_{d}^{\dagger}\left(\tau\right),\label{eq:rho00ideal}
\end{equation}
where $U_{d}\left(\tau\right)$ is given by Eq. (\ref{eq:Ud}) with
$g_{a}=g_{b}=g,$ which corresponds to the \emph{$i{\rm SWAP}$} gate
at the times 
\begin{equation}
\tau_{n}=\left(4n+1\right)\frac{\pi\tilde{\Delta}}{2g^{2}},\ \ \ n=0,1,2,\ldots\label{eq:taun}
\end{equation}

\begin{figure}
\begin{centering}
\includegraphics[bb=0bp 50bp 441bp 293bp,scale=0.52]{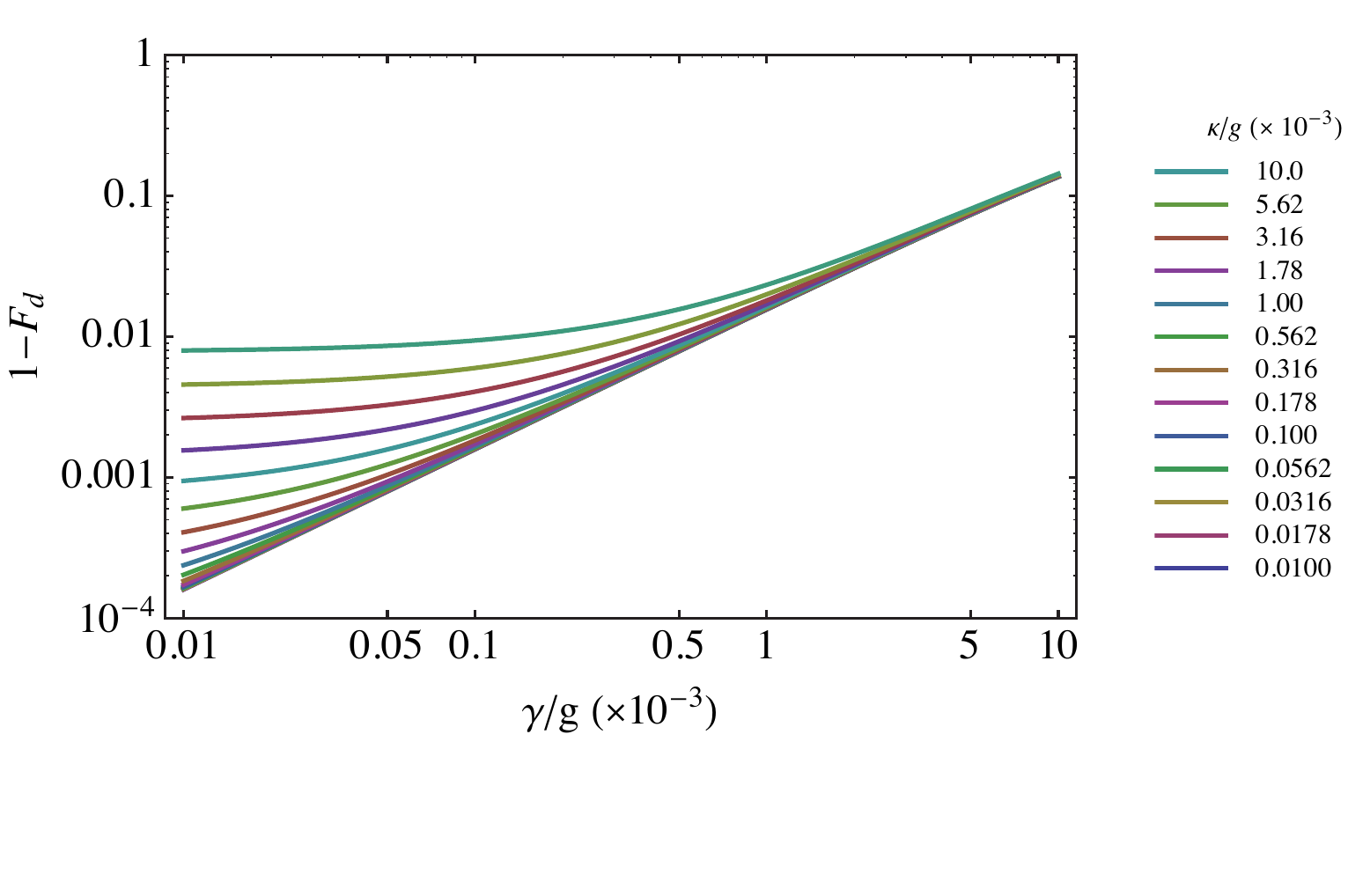}
\par\end{centering}

\protect\caption{\label{fig:Fd}Fidelity $F_{d}\left(\tau_{0}\right)$ of the \emph{$i{\rm SWAP}$}
gate as a function of the qubit decay rate $\gamma$ for several values
of the photon decay rate $\kappa,$ with both rates expressed in units
of the coupling strength $g.$ The fidelity is calculated using Eq.
(\ref{eq:Fd}) for $\tilde{\Delta}/g=20.$ }
\end{figure}

We calculate the \emph{$i{\rm SWAP}$} gate fidelity via 
\begin{equation}
F_{d}\left(\tau_{n}\right)\equiv{\rm Tr}\left[\rho_{00,{\rm id}}\left(\tau_{n}\right)\rho_{00}\left(\tau_{n}\right)\right]\label{eq:Fd}
\end{equation}
for the initial state in Eq. (\ref{eq:rho0}). The gate error $1-F_{d}$
for $\tilde{\Delta}/g=20$ and $n=0$ (i.e., at time $\tau_{0}=\pi\tilde{\Delta}/2g^{2}=10\pi/g$)
is shown in Fig. \ref{fig:Fd} as a function of $\gamma/g$ and $\kappa/g,$
where the chosen ranges of decay rates lie well within the strong
coupling regime described by $\gamma/g<1$ and $\kappa/g<1$. We see
that the error depends more sensitively on $\gamma/g$ than on $\kappa/g,$
as is expected for the dispersive regime of RX qubit-resonator coupling.
For $g=2\pi\times9\ {\rm MHz}$ and $\omega_{0}=2\pi\times1.5\ {\rm GHz}$
(see Sec. \ref{sec:RXrescoupling}), the chosen parameter values correspond
to $\omega=2\pi\times1.3\ {\rm GHz}$ and the $i{\rm SWAP}$ gate
time $\tau_{0}=540\ {\rm ns},$ while gate fidelities greater than
$0.99$ are expected for $\gamma\lesssim2\pi\times0.9\ {\rm kHz}$
(or $T_{2}^{\ast}\gtrsim170\ \mu{\rm s}$) and $\kappa\lesssim2\pi\times90\ {\rm kHz}$
(or $Q\gtrsim1.6\times10^{4}$). These upper bounds on $\gamma$ and
$\kappa$ are consistent with the expectation that gate fidelities
for the dispersive regime are limited more by $\gamma$ than by $\kappa$
and suggest higher fidelities for systems in which the qubit coherence
time is much longer than the resonator photon coherence time. Finally,
we note that a controlled-NOT (CNOT) entangling gate can be constructed
using two $i{\rm SWAP}$ gates combined with single-qubit rotations
\cite{Schuch2003}. As all single-qubit rotations for RX qubits can
be generated via exchange \cite{Medford2013,Taylor2013} and $\tau_{0}\gg\omega^{-1}\sim1\ {\rm ns}$
(see Sec. \ref{sub:resonant}), we can estimate the total CNOT gate
time as $\tau_{{\rm CNOT}}\sim2\tau_{0}\sim1\ \mu{\rm s}.$

\subsection{\label{sub:Fresonant}Driven resonant regime }

We now consider the driven resonant regime discussed in Sec. \ref{sub:resonant},
in which two RX qubits are entangled through sideband transitions
generated by a combination of microwave driving of the resonator and
the individual qubit-resonator interactions. As the sideband transitions
represent the main entangling mechanism involved in the multi-gate
sequence for the controlled-Z gate given in Eq. (\ref{eq:Ucz}), we
focus on a single sideband transition in order to obtain insight into
the dependence of the entangling gate fidelity on the qubit and photon
decay rates for the driven resonant regime. We therefore estimate
the fidelity of a sideband $\pi$ pulse for one RX qubit coupled to
the fundamental mode of the resonator. 

In order to describe the dynamics in the presence of qubit dephasing
and cavity decay with rates $\gamma$ and $\kappa,$ respectively,
we begin with the master equation 

\begin{eqnarray}
\dot{\rho}_{r} & = & -i\left[H_{r},\rho_{r}\right]+\frac{\gamma}{2}\left(\sigma^{z}\rho_{r}\sigma^{z}-\rho_{r}\right)\nonumber \\
 & + & \frac{\kappa}{2}\left(2\tilde{a}\rho_{r}\tilde{a}^{\dagger}-\tilde{a}^{\dagger}\tilde{a}\rho_{r}-\rho_{r}\tilde{a}^{\dagger}\tilde{a}\right),\label{eq:rhormeq}
\end{eqnarray}
with $H_{r}$ given in Eq. (\ref{eq:Hr}). As in Eq. (\ref{eq:Hrprime}),
we apply a displacement transformation $D\left(\alpha^{\prime}\right)\equiv e^{\alpha^{\prime}\tilde{a}^{\dagger}-\alpha^{\prime\ast}\tilde{a}}$
in order to eliminate the direct action of the drive on the resonator,
where $\alpha^{\prime}\left(t\right)=-\varepsilon e^{-i(\nu\tau+\phi)}/\left(\Delta_{0}-i\kappa/2\right)$
is the steady-state solution of $\dot{\alpha}^{\prime}+\left(i\omega_{0}+\kappa/2\right)\alpha^{\prime}+i\varepsilon e^{-i(\nu\tau+\phi)}=0$
and $\Delta_{0}\equiv\omega_{0}-\nu.$ Defining $\rho_{r}^{\prime}\equiv D^{\dagger}\left(\alpha^{\prime}\right)\rho_{r}D\left(\alpha^{\prime}\right),$
we find 
\begin{eqnarray}
\dot{\rho}_{r}^{\prime} & = & -i\left[H_{r}^{\prime\prime},\rho_{r}^{\prime}\right]+\frac{\gamma}{2}\left(\sigma^{z}\rho_{r}^{\prime}\sigma^{z}-\rho_{r}^{\prime}\right)\nonumber \\
 & + & \frac{\kappa}{2}\left(2\tilde{a}\rho_{r}^{\prime}\tilde{a}^{\dagger}-\tilde{a}^{\dagger}\tilde{a}\rho_{r}^{\prime}-\rho_{r}^{\prime}\tilde{a}^{\dagger}\tilde{a}\right),\label{eq:rhorprimemeq}
\end{eqnarray}
where
\begin{equation}
H_{r}^{\prime\prime}=\omega_{0}\tilde{a}^{\dagger}\tilde{a}+\frac{\omega}{2}\sigma^{z}+g\sigma^{x}\left(\tilde{a}^{\dagger}+\tilde{a}\right)-2\Omega^{\prime}\cos\left(\nu\tau\right)\sigma^{x}.\label{eq:Hrdbprime}
\end{equation}
Here, the Rabi frequency is given by 
\begin{equation}
2\Omega^{\prime}\equiv\frac{2g\varepsilon}{\sqrt{\Delta_{0}^{2}+\kappa^{2}/4}},\label{eq:Rabifreqprime}
\end{equation}
where we include the phase $\phi$ of the driving field in the definition
of $\Omega^{\prime}$ and set $\tan\phi=\kappa/2\Delta_{0},$ such
that $\Omega^{\prime}$ is real even for $\kappa\neq0.$ 

We now transform the master equation to the frame rotating at the
drive frequency $\nu.$ By applying Eq. (\ref{eq:U1}) to the density
matrix such that $\rho_{r}^{{\rm rf}}\equiv U_{1}^{\dagger}\rho_{r}^{\prime}U_{1}$
and setting $\nu=\omega,$ we can rewrite Eq. (\ref{eq:rhorprimemeq})
as 
\begin{eqnarray}
\dot{\rho}_{r}^{{\rm rf}} & = & -i\left[H_{r}^{{\rm rf\prime}},\rho_{r}^{{\rm rf}}\right]+\frac{\gamma}{2}\left(\sigma^{z}\rho_{r}^{{\rm rf}}\sigma^{z}-\rho_{r}^{{\rm rf}}\right)\nonumber \\
 & + & \frac{\kappa}{2}\left(2\tilde{a}\rho_{r}^{{\rm rf}}\tilde{a}^{\dagger}-\tilde{a}^{\dagger}\tilde{a}\rho_{r}^{{\rm rf}}-\rho_{r}^{{\rm rf}}\tilde{a}^{\dagger}\tilde{a}\right),\label{eq:rhorrfmeq}
\end{eqnarray}
where 
\begin{equation}
H_{r}^{{\rm rf\prime}}=\Delta_{0}\tilde{a}^{\dagger}\tilde{a}+g\left(\sigma^{+}\tilde{a}+\sigma^{-}\tilde{a}^{\dagger}\right)-\Omega^{\prime}\sigma^{x}\label{eq:Hrrfprime}
\end{equation}
and we have again dropped rapidly oscillating terms $\sim e^{\pm2i\nu\tau}$
as in Eq. (\ref{eq:Hrrf}). Finally, applying the rotations $\tilde{U}_{{\rm rot}}=e^{-i\left(\pi/4\right)\sigma^{z}}$
and $U_{{\rm rot}}^{\prime}$ {[}Eq. (\ref{eq:Urotprime}){]} successively
such that $\tilde{\rho}_{r}^{{\rm rot}}\equiv U_{{\rm rot}}^{\prime\dagger}\tilde{U}_{{\rm rot}}^{\dagger}\rho_{r}^{{\rm rf}}\tilde{U}_{{\rm rot}}U_{{\rm rot}}^{\prime}$
and letting $a\equiv i\tilde{a}$ leads to
\begin{eqnarray}
\dot{\tilde{\rho}}_{r}^{{\rm rot}} & = & -i\left[\tilde{H}_{r}^{{\rm rot}},\tilde{\rho}_{r}^{{\rm rot}}\right]+\frac{\gamma}{2}\left(\sigma^{y}\tilde{\rho}_{r}^{{\rm rot}}\sigma^{y}-\tilde{\rho}_{r}^{{\rm rot}}\right)\nonumber \\
 & + & \frac{\kappa}{2}\left(2a\tilde{\rho}_{r}^{{\rm rot}}a^{\dagger}-a^{\dagger}a\tilde{\rho}_{r}^{{\rm rot}}-\tilde{\rho}_{r}^{{\rm rot}}a^{\dagger}a\right),\label{eq:rhorrottildemeq}
\end{eqnarray}
where 
\begin{eqnarray}
\tilde{H}_{r}^{{\rm rot}} & = & \Delta_{0}a^{\dagger}a+\Omega^{\prime}\sigma^{z}+\frac{g}{2}\left(\sigma^{+}a+\sigma^{-}a^{\dagger}+\sigma^{+}a^{\dagger}+\sigma^{-}a\right)\nonumber \\
 & + & i\frac{g}{2}\sigma^{z}\left(a-a^{\dagger}\right).\label{eq:Hrrottilde}
\end{eqnarray}
For $\Delta_{0}=2\Omega^{\prime},$ the time evolution generated by
$\tilde{H}_{r}^{{\rm rot}}$ closely approximates the sideband transitions
obtained in the doubly rotating frame {[}compare the third term in
Eq. (\ref{eq:Hrrottilde}) with Eqs. (\ref{eq:Hminus}) and (\ref{eq:Hplus}){]},
as we describe below. For simplicity, we therefore calculate the fidelity
with respect to the ideal density matrix evolution generated by Eq.
(\ref{eq:Hrrottilde}). 

Confining our description to the resonator photon subspaces with $n=0,1,2,$
we choose the initial qubit-resonator state
\begin{equation}
\tilde{\rho}_{r}^{{\rm rot}}\left(0\right)=\ket{e,0}\bra{e,0}.\label{eq:rhorrottilde0}
\end{equation}
The ideal final state generated by a red sideband $\pi$ pulse $S^{-}\left(\pi,0\right)$
{[}see (\ref{eq:Sminus}){]} is then given by 
\begin{eqnarray}
\rho_{\pi} & \equiv & S^{-}\left(\pi,0\right)\tilde{\rho}_{r}^{{\rm rot}}\left(0\right)S^{-\dagger}\left(\pi,0\right)\label{eq:rhopi}\\
 & = & \ket{g,1}\bra{g,1}\nonumber 
\end{eqnarray}
Comparing this state with the final state $\tilde{\rho}_{r}^{{\rm rot,id}}\left(\tau_{{\rm \pi}}\right)$
after ideal evolution for a time $\tau_{{\rm \pi}}=\pi/g$ under the
Hamiltonian in Eq. (\ref{eq:Hrrottilde}) with $\Delta_{0}=2\Omega^{\prime}$
yields ${\rm Tr}\left[\rho_{\pi}\tilde{\rho}_{r}^{{\rm rot,id}}\left(\tau_{{\rm \pi}}\right)\right]\approx0.996.$
Thus, the ideal evolution generated by $\tilde{H}_{r}^{{\rm rot}}$
can itself be regarded approximately as the red sideband transition
$S^{-}\left(\pi,0\right).$ 

\begin{figure}
\begin{centering}
\includegraphics[bb=0bp 50bp 441bp 293bp,scale=0.52]{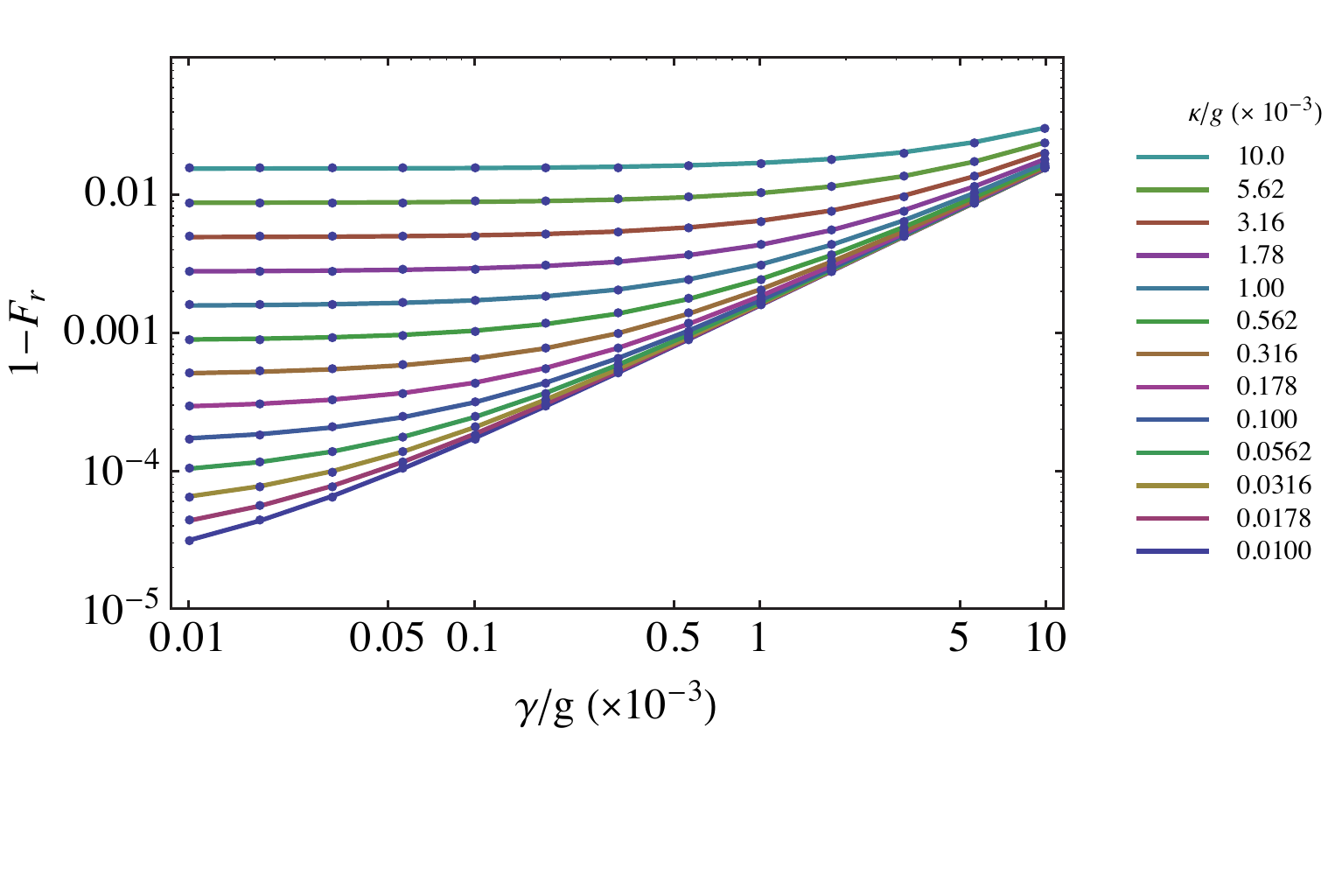}
\par\end{centering}

\protect\caption{\label{fig:Fr}Fidelity $F_{r}\left(\tau_{{\rm \pi}}\right)$ of the
evolution generated by $\tilde{H}_{r}^{{\rm rot}}$ for $\tau_{\pi}=\pi/g,$
corresponding approximately to a sideband $\pi$ pulse, as a function
of the qubit and photon decay rates $\gamma$ and $\kappa,$ respectively,
in units of the coupling strength $g,$ calculated using Eq. (\ref{eq:Fr})
for $\Delta_{0}/g=20.$ Lines are guides for the eye.}
\end{figure}

We numerically calculate the fidelity 
\begin{equation}
F_{r}\left(\tau_{{\rm \pi}}\right)\equiv{\rm Tr}\left[\tilde{\rho}_{r}^{{\rm rot,id}}\left(\tau_{{\rm \pi}}\right)\tilde{\rho}_{r}^{{\rm rot}}\left(\tau_{{\rm \pi}}\right)\right]\label{eq:Fr}
\end{equation}
as a function of $\gamma/g$ and $\kappa/g.$ Fig. \ref{fig:Fr} shows
the corresponding error $1-F_{r}$ for $\Delta_{0}/g=20.$ We see
that, compared to the $i{\rm SWAP}$ gate in the dispersive regime
(Fig. \ref{fig:Fd}), the fidelity for the sideband $\pi$ pulse in
the driven resonant regime depends more sensitively on the photon
decay rate $\kappa/g,$ while the maximum error is approximately one
order of magnitude smaller over the same range of $\gamma/g$ and
$\kappa/g.$ The parameter values $\Delta_{0}/g=20,$ $g=2\pi\times9\ {\rm MHz},$
and $\omega_{0}=2\pi\times1.5\ {\rm GHz}$ lead to $\omega=2\pi\times1.3\ {\rm GHz}$
and the gate time $\tau_{\pi}=54\ {\rm ns},$ which is a factor of
ten shorter than the $i{\rm SWAP}$ gate time $\tau_{0}$ found in
Sec. \ref{sub:Fdisp}. From Fig. \ref{fig:Fr}, gate fidelities greater
than $0.99$ are expected for $\gamma\lesssim2\pi\times9\ {\rm kHz}$
(or $T_{2}^{\ast}\gtrsim17\ \mu{\rm s}$) and $\kappa\lesssim2\pi\times51\ {\rm kHz}$
(or $Q\gtrsim2.9\times10^{4}$). Note that the minimum value of $T_{2}^{\ast}$
is also reduced by a factor of ten compared to that for the dispersive
regime found in Sec. \ref{sub:Fdisp}, suggesting that (provided resonators
with sufficiently high quality factors are available) sideband-based
entangling gates in the driven resonant regime may prove more advantageous
for implementations. In addition, an estimate of the controlled-Z
gate time based on the five sideband pulses appearing in Eq. (\ref{eq:Ucz})
yields $\tau_{{\rm CZ}}\sim5\pi/g\approx270\ \mu{\rm s}.$ As in the
previous section, we again neglect the rapid single-qubit exchange
rotations (see also Sec. \ref{sub:resonant}), which also serve to
convert between the controlled-Z and CNOT gates. Thus, although a
controlled-Z gate in the driven resonant regime requires more sideband
pulses compared to the two $i{\rm SWAP}$ gates needed for a CNOT
gate in the dispersive regime, we nevertheless find that the total
gate time $\tau_{{\rm CZ}}\ll\tau_{{\rm CNOT}}.$

\subsection{\label{sub:Fdrivdisp}Driven dispersive regime}

Finally, we consider the gate fidelity for the driven dispersive regime
(Sec. \ref{sub:drivdisp}), in which both RX qubits are again coupled
to the resonator in the dispersive regime (see Sec. \ref{sub:Fdisp}),
but with $\omega_{a}\neq\omega_{b}$ and a microwave driving field
applied to the resonator {[}see Sec. \ref{sub:Fresonant} and the
last term of Eq. (\ref{eq:Hr}){]}. The master equation in the presence
of qubit and cavity decay has the form 
\begin{eqnarray}
\dot{\rho}_{dd} & = & -i\left[H_{dd},\rho_{dd}\right]+\sum_{\mu=a,b}\frac{\gamma_{\mu}}{2}\left(\sigma_{\mu}^{z}\rho_{dd}\sigma_{\mu}^{z}-\rho_{dd}\right)\nonumber \\
 & + & \frac{\kappa}{2}\left(2a\rho_{dd}a^{\dagger}-a^{\dagger}a\rho_{dd}-\rho_{dd}a^{\dagger}a\right),\label{eq:rhoddmeq}
\end{eqnarray}
where $H_{dd}$ is given in Eq. (\ref{eq:Hdd}). As in Sec. \ref{sub:Fresonant}
and Eq. (\ref{eq:Hddprime}), we eliminate the direct action of the
driving field on the resonator via a displacement transformation $D\left(\alpha^{\prime}\right)\equiv e^{\alpha^{\prime}a^{\dagger}-\alpha^{\prime\ast}a}.$
Re-expressing the master equation {[}Eq. (\ref{eq:rhoddmeq}){]} in
terms of $\rho_{dd}^{\prime}\equiv D^{\dagger}\left(\alpha^{\prime}\right)\rho_{dd}D\left(\alpha^{\prime}\right)$
leads to 
\begin{eqnarray}
\dot{\rho}_{dd}^{\prime} & = & -i\left[H_{dd}^{\prime\prime},\rho_{dd}^{\prime}\right]+\sum_{\mu=a,b}\frac{\gamma_{\mu}}{2}\left(\sigma_{\mu}^{z}\rho_{dd}^{\prime}\sigma_{\mu}^{z}-\rho_{dd}^{\prime}\right)\nonumber \\
 & + & \frac{\kappa}{2}\left(2a\rho_{dd}^{\prime}a^{\dagger}-a^{\dagger}a\rho_{dd}^{\prime}-\rho_{dd}^{\prime}a^{\dagger}a\right),\label{eq:rhoddprimemeq}
\end{eqnarray}
where
\begin{eqnarray}
H_{dd}^{\prime\prime} & = & \omega_{0}a^{\dagger}a+\sum_{\mu=a,b}\frac{\omega_{\mu}}{2}\sigma_{\mu}^{z}+\sum_{\mu=a,b}g_{\mu}\sigma_{\mu}^{x}\left(a+a^{\dagger}\right)\nonumber \\
 &  & -\sum_{\mu=a,b}2\Omega_{\mu}^{\prime}\cos\left(\nu t\right)\sigma_{\mu}^{x}\label{eq:Hdddbprime}
\end{eqnarray}
with 
\begin{equation}
2\Omega_{\mu}^{\prime}=\frac{2g_{\mu}\varepsilon}{\sqrt{\Delta_{0}^{2}+\kappa^{2}/4}}.\label{eq:Rabifreqmuprime}
\end{equation}
As in Eq. (\ref{eq:Rabifreqprime}), we include the phase $\phi$
in the Rabi frequencies and choose $\phi$ to satisfy $\tan\phi=\kappa/2\Delta_{0}$
such that $\Omega_{\mu}^{\prime}$ is real for $\mu=a,b.$ 

Transforming the master equation to a rotating frame via Eq. (\ref{eq:U1prime}),
defining $\rho_{dd}^{{\rm rf}}\equiv U_{1}^{\prime\dagger}\rho_{dd}^{\prime}U_{1}^{\prime},$
assuming that qubit $a$ is driven resonantly such that $\nu=\omega_{a},$
and setting $\Omega_{b}^{\prime}=0$ (see Sec. \ref{sub:drivdisp})
gives
\begin{eqnarray}
\dot{\rho}_{dd}^{{\rm rf}} & = & -i\left[H_{dd}^{{\rm rf}\prime},\rho_{dd}^{{\rm rf}}\right]+\sum_{\mu=a,b}\frac{\gamma_{\mu}}{2}\left(\sigma_{\mu}^{z}\rho_{dd}^{{\rm rf}}\sigma_{\mu}^{z}-\rho_{dd}^{{\rm rf}}\right)\nonumber \\
 & + & \frac{\kappa}{2}\left(2a\rho_{dd}^{{\rm rf}}a^{\dagger}-a^{\dagger}a\rho_{dd}^{{\rm rf}}-\rho_{dd}^{{\rm rf}}a^{\dagger}a\right),\label{eq:rhoddrfmeq}
\end{eqnarray}
where 
\begin{eqnarray}
H_{dd}^{{\rm rf\prime}} & = & \Delta_{0}a^{\dagger}a-\Omega_{a}^{\prime}\sigma_{a}^{x}+\frac{\delta}{2}\sigma_{b}^{z}+\sum_{\mu=a,b}g_{\mu}\left(\sigma_{\mu}^{+}a+\sigma_{\mu}^{-}a^{\dagger}\right),\nonumber \\
\label{eq:Hddrfprime}
\end{eqnarray}
and rapidly oscillating terms $\sim e^{\pm2i\nu\tau}$ have been dropped.
Finally, applying the rotation in Eq. (\ref{eq:Ua}) leads to 
\begin{eqnarray}
\dot{\rho}_{dd}^{{\rm rot}} & = & -i\left[H_{dd}^{{\rm rot\prime}},\rho_{dd}^{{\rm rot}}\right]+\frac{\gamma_{a}}{2}\left(\sigma_{a}^{x}\rho_{dd}^{{\rm rot}}\sigma_{a}^{x}-\rho_{dd}^{{\rm rot}}\right)\nonumber \\
 & + & \frac{\gamma_{b}}{2}\left(\sigma_{b}^{z}\rho_{dd}^{{\rm rot}}\sigma_{b}^{z}-\rho_{dd}^{{\rm rot}}\right)\nonumber \\
 & + & \frac{\kappa}{2}\left(2a\rho_{dd}^{{\rm rot}}a^{\dagger}-a^{\dagger}a\rho_{dd}^{{\rm rot}}-\rho_{dd}^{{\rm rot}}a^{\dagger}a\right),\label{eq:rhoddrotmeq}
\end{eqnarray}
where $H_{dd}^{{\rm rot\prime}}=H_{0}^{\prime\prime}+V_{dd},$ with
\begin{eqnarray}
H_{0}^{\prime\prime} & = & \Delta_{0}a^{\dagger}a+\Omega_{a}^{\prime}\sigma_{a}^{z}+\frac{\delta}{2}\sigma_{b}^{z}\label{eq:H0dbprime}
\end{eqnarray}
and $V_{dd}$ given by Eq. (\ref{eq:Vdd}). 

We now follow an approach similar to that used for the dispersive
regime in Sec. \ref{sub:Fdisp} in order to numerically calculate
the fidelity using Eq. (\ref{eq:rhoddrotmeq}). Confining the description
to the $n=0$ and $n=1$ subspaces, we find the subspace projections
of $H_{dd}^{{\rm rot}\prime}$
\begin{eqnarray}
h_{0}^{\prime} & \equiv & PH_{dd}^{{\rm rot}\prime}P=PH_{0}^{\prime\prime}P=\Omega_{a}^{\prime}\sigma_{a}^{z}+\frac{\delta}{2}\sigma_{b}^{z},\label{eq:h0prime}\\
h_{1}^{\prime} & \equiv & QH_{dd}^{{\rm rot}\prime}Q=QH_{0}^{\prime\prime}Q=\Delta_{0}+\Omega_{a}^{\prime}\sigma_{a}^{z}+\frac{\delta}{2}\sigma_{b}^{z},\label{eq:h1prime}\\
v^{\prime} & \equiv & PH_{dd}^{{\rm rot}\prime}Q=PV_{dd}Q=\frac{g_{a}}{2}\left(-\sigma_{a}^{z}+i\sigma_{a}^{y}\right)+g_{b}\sigma_{b}^{+}.\nonumber \\
\label{eq:vprime}
\end{eqnarray}
In terms of the associated subspace projections of the density matrix
$\rho_{00}^{\prime}\equiv P\rho_{dd}^{{\rm rot}}P,$ $\rho_{01}^{\prime}\equiv P\rho_{dd}^{{\rm rot}}Q$
and $\rho_{11}^{\prime}=0$ (see Sec. \ref{sub:Fdisp}), Eq. (\ref{eq:rhoddrotmeq})
yields 
\begin{eqnarray}
\dot{\rho}_{00}^{\prime} & = & -i\left(\left[h_{0}^{\prime},\rho_{00}^{\prime}\right]+v^{\prime}\rho_{01}^{\prime\dagger}-\rho_{01}^{\prime}v^{\prime\dagger}\right)\nonumber \\
 & + & \frac{\gamma_{a}}{2}\left(\sigma_{a}^{x}\rho_{00}^{\prime}\sigma_{a}^{x}-\rho_{00}^{\prime}\right)+\frac{\gamma_{b}}{2}\left(\sigma_{b}^{z}\rho_{00}^{\prime}\sigma_{b}^{z}-\rho_{00}^{\prime}\right),\nonumber \\
\label{eq:rho00primemeq}\\
\dot{\rho}_{01}^{\prime} & = & -i\left(h_{0}^{\prime}\rho_{01}^{\prime}-\rho_{01}^{\prime}h_{1}^{\prime}-\rho_{00}^{\prime}v^{\prime}\right)\nonumber \\
 & + & \frac{\gamma_{a}}{2}\left(\sigma_{a}^{x}\rho_{01}^{\prime}\sigma_{a}^{x}-\rho_{01}^{\prime}\right)+\frac{\gamma_{b}}{2}\left(\sigma_{b}^{z}\rho_{01}^{\prime}\sigma_{b}^{z}-\rho_{01}^{\prime}\right)\nonumber \\
 & - & \frac{\kappa}{2}\rho_{01}^{\prime}.\label{eq:rho01primemeq}
\end{eqnarray}
As described in Sec. \ref{sub:Fdisp}, we set $\dot{\rho}_{01}^{\prime}=0$
and apply a Liouville-space mapping in order to solve for $\rho_{00}^{\prime}\left(\tau\right).$ 

In Sec. \ref{sub:drivdisp}, we showed that the $i{\rm SWAP}$ entangling
gate can be generated by the interaction in a doubly rotating frame
for $2\tilde{\Omega}_{a}=\tilde{\delta}$ {[}see Eq. (\ref{eq:Hdddrf}){]}.
Setting $2\Omega_{a}^{\prime}=\delta,$ we incorporate this condition
into the present analysis via Eq. (\ref{eq:gconstraint}). We again
take $\rho_{00}\left(0\right)$ {[}Eq. (\ref{eq:rho0}){]} as the
initial state of the system and obtain $\rho_{00}^{\prime}\left(\tau\right)$
numerically. Defining $g\equiv g_{a},$ setting $\Delta_{0}/g=20,$
and choosing the $i{\rm SWAP}$ gate time 
\begin{eqnarray}
\tau_{0} & = & \pi\frac{\Delta_{0}-\delta}{2g^{2}}\sqrt{\frac{2\left(\Delta_{0}+\delta\right)}{\delta}}\nonumber \\
 & = & 32\sqrt{3}\frac{\pi}{g},\label{eq:tau0}
\end{eqnarray}
we find ${\rm Tr}\left[\rho_{00,{\rm id}}\left(\tau_{0}\right)\rho_{00,{\rm id}}^{\prime}\left(\tau_{0}\right)\right]\approx0.999,$
indicating that the evolution generated by $H_{dd}^{{\rm rot\prime}}$
for $2\Omega_{a}^{\prime}=\delta$ and the constraint in Eq. (\ref{eq:gconstraint})
closely matches an ideal $i{\rm SWAP}$ gate. Thus, we approximate
the $i{\rm SWAP}$ gate fidelity for the driven dispersive regime
by 
\begin{equation}
F_{dd}\left(\tau_{0}\right)\equiv{\rm Tr}\left[\rho_{00,{\rm id}}^{\prime}\left(\tau_{0}\right)\rho_{00}^{\prime}\left(\tau_{0}\right)\right].\label{eq:Fdd}
\end{equation}

\begin{figure}
\begin{centering}
\includegraphics[bb=0bp 50bp 440bp 295bp,scale=0.52]{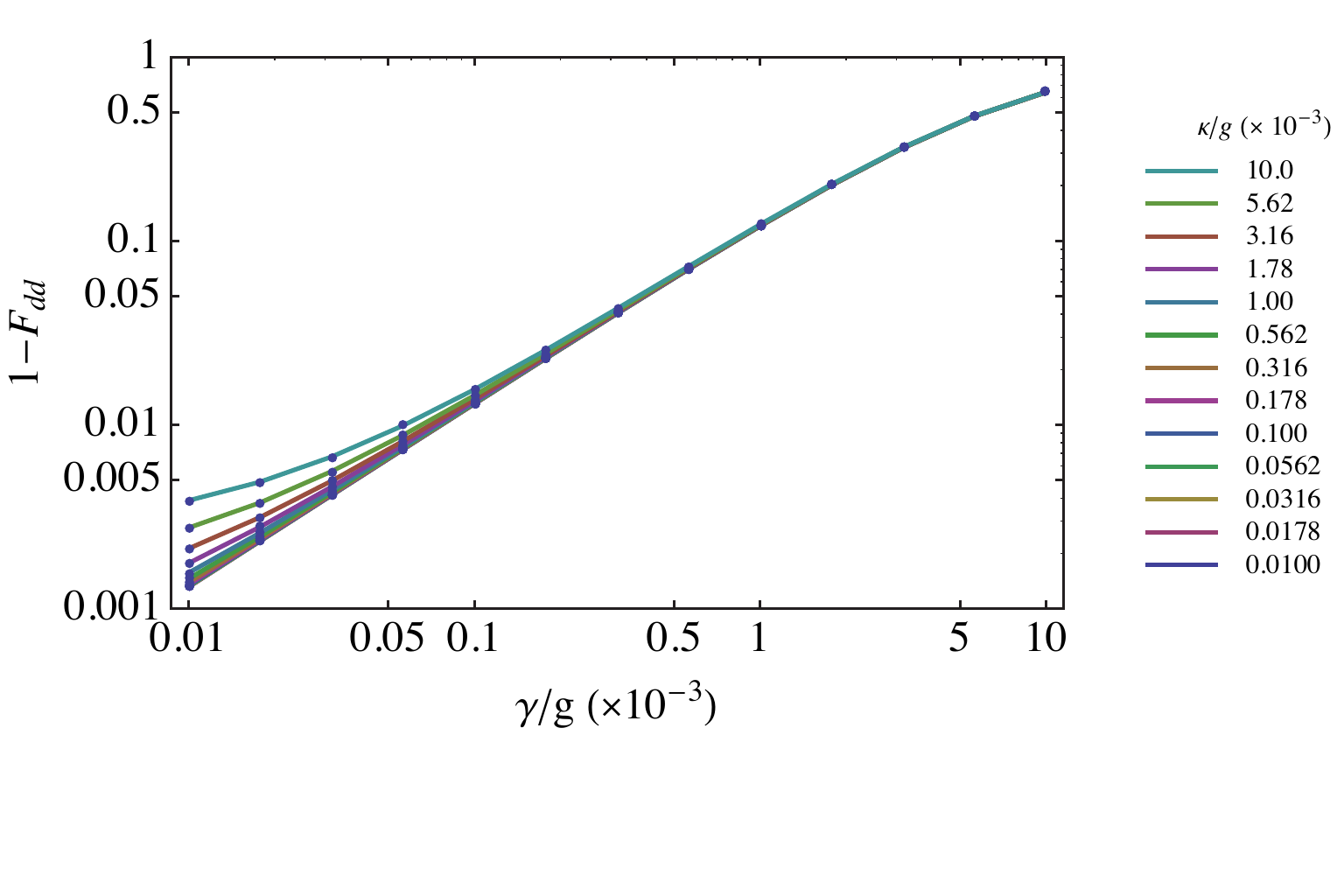}
\par\end{centering}

\protect\caption{\label{fig:Fdd}Fidelity $F_{dd}\left(\tau_{0}\right)$ of the evolution
generated by $H_{dd}^{{\rm rot\prime}}$ as a function of the qubit
and photon decay rates $\gamma$ and $\kappa,$ respectively, in units
of the coupling strength $g,$ calculated using Eq. (\ref{fig:Fdd})
for $\Delta_{0}/g=20$ and $\delta/g=4.$ Lines are guides for the
eye. }
\end{figure}

Figure \ref{fig:Fdd} shows the error $1-F_{dd}$ as a function of
the qubit and cavity decay rates $\gamma/g$ and $\kappa/g$ for $\Delta_{0}/g=20$
and $\delta/g=4.$ We note that, as in the dispersive regime considered
in Sec. \ref{sub:Fdisp}, the error depends more sensitively on $\gamma/g$
than $\kappa/g$ and is also larger than that for the driven resonant
regime. For $\Delta_{0}/g=20,$ $g=2\pi\times9\ {\rm MHz,}$ and $\omega_{0}=2\pi\times1.5\ {\rm GHz,}$
we find $\omega=2\pi\times1.3\ {\rm GHz}$ and the $i{\rm SWAP}$
gate time $\tau_{0}=3.0\ \mu{\rm s}.$ Gate fidelities greater than
$0.99$ correspond to $\gamma\lesssim2\pi\times0.51\ {\rm kHz}$ (or
$T_{2}^{\ast}\gtrsim310\ \mu{\rm s}$) and $\kappa\lesssim2\pi\times90\ {\rm kHz}$
(or $Q\gtrsim1.6\times10^{4}$). While the longer gate time $\tau_{0}$
relative to that chosen in Sec. \ref{sub:Fdisp} leads to a longer
minimum $T_{2}^{\ast},$ we again find that the gate fidelity should
be higher for systems in which the coherence time of the qubit is
much longer than that of the resonator photons, as is expected for
the dispersive regime. Finally, we use the approach described in Sec.
\ref{sub:Fdisp} to estimate the total CNOT gate time as $\tau_{{\rm CNOT}}\sim2\tau_{0}\sim6\ \mu{\rm s}$
and thus again find that $\tau_{{\rm CZ}}\ll\tau_{{\rm CNOT}}$ (see
Sec. \ref{sub:Fresonant}).

\section{Conclusions}

In the present work, we have analyzed three approaches drawn from
a combination of circuit QED and Hartmann-Hahn double resonance techniques
for entangling spatially separated RX qubits via a superconducting
transmission line resonator. We derived both the form of the RX qubit-resonator
coupling and resonator-mediated entangling gates in the dispersive,
driven resonant, and driven dispersive regimes. While both dispersive
regimes yield two-qubit gate rates $\sim\xi^{4},$ where $\xi\equiv t/\Delta$
is the charge admixture parameter for the RX qubit and is related
to the qubit-resonator coupling strength via $g\sim\xi^{2}$, rapid
gates with rates $\sim\xi^{2}$ and smaller error are possible in
the driven resonant regime. 

Our results show that an implementation of RX qubits in silicon triple
quantum dots in principle enables robustness to phonon-induced relaxation
and possesses characteristics highly favorable for achieving the strong
coupling regime of interaction with the resonator. Furthermore, the
analysis of gate fidelities for the three regimes we consider suggests
that, while the requirements for resonator quality factors are somewhat
relaxed in the dispersive regimes, high-fidelity entangling gates
based on sideband transitions in the driven resonant regime in combination
with rapid $\left(\tau_{{\rm gate}}\sim\omega^{-1}\lesssim1\ {\rm ns}\right)$
single-qubit rotations via exchange may prove advantageous for implementations
with resonators of sufficiently high quality factors. In particular,
although five sideband pulses are required to construct a controlled-Z
gate according to Eq. (\ref{eq:Ucz}), we expect that the the total
gate time should still be considerably shorter than those for controlled-NOT
gates carried out using two $i{\rm SWAP}$ gates in the dispersive
and driven dispersive regimes. We therefore find that the exchange-based
universal control intrinsic to RX qubits enables rapid entangling
gates in the driven resonant regime.

Many potential future directions remain to be explored. Experiments
will ultimately provide more insight into the achievable coherence
times, resonator quality factors, and optimal coupling regime for
entanglement within the RX qubit-resonator system. Identifying methods
for integrating this basic unit into a robust modular architecture
also remains an open challenge. Additionally, while the focus of the
present work is on so-called transverse RX qubit-resonator dipole
coupling of the form $\sigma_{x}\left(a+a^{\dagger}\right),$ future
work may involve investigating potential improvements in the performance
of entangling gates for RX qubits via longitudinal dipole coupling
of the form $\sigma_{z}\left(a+a^{\dagger}\right)$ \cite{Jin2012,Billangeon2015}.
Finally, the implementation of RX qubits in silicon will provide an
opportunity to verify the expected improvements in coherence compared
to RX qubits in GaAs.
\begin{acknowledgments}
We thank B. E. Kane and F. W. Strauch for helpful discussions. 
\end{acknowledgments}

\bibliographystyle{apsrev4-1}
\bibliography{RXres}

\end{document}